\documentclass{aastex}
\usepackage{emulateapj5}
\usepackage{graphicx}
\usepackage{natbib}
\newcommand{\el}[2]{\ensuremath{^{#1}\mathrm{#2}}}
\newcommand{\Mo}{\rm{M}_\odot}

\def\mnras{MNRAS}
\def\araa{ARA\&A}
\def\apj{ApJ}
\def\aap{A\&A}
\def\aj{AJ}
\def\pasp{PASP}
\def\apjl{ApJ}
\def\apjs{ApJS}
\def\memsai{Mem. S. A. It}
\begin{document}
\shorttitle{Thermohaline in GCs}
\shortauthors{Angelou et al.}

\title{The Role of Thermohaline Mixing in Intermediate- and Low-Metallicity
Globular Clusters}
\author{George C. Angelou\altaffilmark{1}, Richard J.
Stancliffe\altaffilmark{2,1},   Ross P.
Church\altaffilmark{3,1}, John C.
Lattanzio\altaffilmark{1}, Graeme H. Smith\altaffilmark{4} .}
\altaffiltext{1}{Monash Centre for Astrophysics,  School of
Mathematical Sciences, Monash University,  Melbourne,  VIC 3800,  Australia.}
\altaffiltext{2}{Research School of Astronomy \& Astrophysics, Mt Stromlo
Observatory, Canberra, ACT 2611, Australia.}
\altaffiltext{3}{Department of Astronomy and Theoretical Physics, Lund
Observatory, Box 43, SE-221 00 Lund, Sweden.}
\altaffiltext{4}{University of California Observatories/Lick Observatory,
Department of Astronomy and Astrophysics, UC Santa Cruz,1156 High St., Santa
Cruz, CA 95064, USA.}

\email{George.Angelou@monash.edu}

\begin{abstract}
It is now widely accepted that globular cluster red giant branch 
stars owe their strange abundance patterns to a combination of pollution from
progenitor stars and in situ extra mixing. In this hybrid theory a first
generation of stars imprint abundance patterns  into
the gas from which a second
generation forms. The hybrid theory suggests that extra mixing is operating in
both  populations and we use the variation of [C/Fe] with luminosity to examine
how efficient this mixing is.  We investigate the observed red giant branches
of M3, M13,
M92, M15 and
NGC 5466 as a means to test a theory of thermohaline mixing. The second
parameter pair M3 and M13 are of intermediate metallicity and
our models are able to account for the evolution of carbon along the RGB in both
clusters.   Although, in order to fit the most carbon-depleted main-sequence
stars in M13 we require
a model whose initial [C/Fe] abundance leads to a carbon abundance lower than
is
observed.
Furthermore our results suggest that
stars in M13 formed with some primary nitrogen (higher C+N+O than stars in M3).
In the metal-poor
regime only NGC 5466 can be tentatively explained by thermohaline mixing
operating in multiple populations.  We find thermohaline mixing unable
to model the depletion of [C/Fe] with magnitude in M92 and M15.  
It appears as if extra mixing is occurring before the luminosity
function bump in these clusters. To reconcile the data with the models would
require first
dredge-up to be deeper than found in extant models.

\end{abstract}


%%%%%%%%%%%   Introduction  %%%%%%%%%%%%%%%%
\section{Introduction} \label{intro}
It is clear from the annals of stellar astrophysics that low-mass 
stars undergo mixing as they ascend the red giant branch (RGB hereafter, for
example see
\citealt{1998A&A...336..915C}). This is contrary to canonical models of red
giants that incorporate convection as the only form of internal mass transport.
Such models predict just one event capable of changing surface abundances of
light elements on the first ascent of the red giant branch, namely the
so-called first dredge up (FDU e.g., \citealt{1967ApJ...147..624I}). However, a
variety of observations indicate
that some form of non-convective mixing (which is often termed ``extra mixing"
in the literature) progressively transports CN-processed material to the
surface along the upper RGB whilst reducing the \el{12}{C} abundance.
 The effects of extra mixing are observed in field
stars
\citep{2006A&A...455..291S,2005A&A...430..655S,2000A&A...354..169G,
1998A&A...332..204Cgr, 1986ApJ...311..826S} and globular cluster (GC) stars
\citep{2003ApJ...585L..45S,2007A&A...461L..13R} with increasing efficiency seen
at lower metallicities.  The mixing occurs internally as the star ascends the
giant branch therefore variations in abundance  manifest themselves as a
function of luminosity. This is most clearly seen in carbon
\citep{2003PASP..115.1211S} and
lithium \citep{2009A&A...503..545L, 2011MNRAS.412...81M}. 

During the RGB phase of evolution the star possesses a
degenerate hydrogen-exhausted core. A thin hydrogen-burning shell progresses
outwards
through the star contributing processed material to the inert core. A convective
envelope which transports material to the stellar surface is separated from the
shell by a radiative region. Extra mixing requires some mechanism to transport
material across
this radiative zone to the top of the hydrogen shell where it
encounters temperatures where burning can occur. Processed material is cycled
back to the convective envelope accounting for the observed compositions.
Identifying the exact physics that induces this mixing has been debated at
length; it has proven to be a challenging task for
stellar astrophysics. 

Phenomenological models of
extra mixing can match the
general
abundance trends associated with further dredge up of CN processed material
\citep{2011ApJ...729....3P, 2011ApJ...741...26P, 2003ApJ...593..509D, 1995ApJ...447L..37W}. Many
physically based mechanisms have also been explored as potential candidates:
rotational mixing
\citep{1979ApJ...229..624S, 2005ApJ...631..540C, 2006A&A...453..261P}, 
magnetic fields \citep{2009PASA...26..161P, 2008ApJ...684L..29N,
2007ApJ...671..802B,1980ApJ...239..248H}, 
internal gravity waves \citep{2000MNRAS.316..395D} and more recently
thermohaline mixing
\citep{2006Sci...314.1580E, 2007A&A...467L..15C, 2008ApJ...677..581E} and the combination of thermohaline mixing and magnetic fields \citep{2009ApJ...696.1823D}. Like any
new
paradigm thermohaline mixing has stimulated a wealth of subsequent work (see
\citealt{2010MNRAS.403..505S, 2010A&A...521A...9C,2010A&A...522A..10C,
2011ApJ...728...79A}). 
The mechanism is  initially driven by instabilities brought
about from local inversions in the mean molecular weight. These inversions
arise through the burning of \el{3}{He}. The reaction $\el{3}{He}\left(
\el{3}{He},2 \rm{p}\right) \el{4}{He}$ produces
more particles than it consumes. If this reaction can occur in a homogenized
region, such as is the case when the hydrogen burning shell advances through the
composition discontinuity left by first dredge-up (FDU), then its effect on the
molecular
weight profile can dominate (for a detailed explanation of the mixing process
see \citealt{2011ApJ...728...79A}). 
Understanding and modelling this mechanism is an area of active research at
present
\citep{2011arXiv1104.0832W, 2011ApJ...728L..29T,
2011ApJ...727L...8D, 2007A&A...476L..29C}.

\citet{2007A&A...467L..15C}
have shown that the mechanism can match the RGB abundances of field stars, with
a dependence on metallicty which was also demonstrated by
\citet{ 2008ApJ...677..581E}. The next logical step
was to
test the mechanism against abundance measurements in globular clusters as  
\citet{2011ApJ...728...79A} did with stars in M3.  M3 was chosen as 
a test case because it is considered a typical globular cluster. A metallicity
of [Fe/H] $= -1.4$ \citep{2004AJ....127.2162S, 
2005AJ....129..303C}
means that it falls near the mode in the metallicity distribution of halo
globular clusters. It is well studied and exhibits 
significant [C/Fe]  depletion \citep{1981ApJS...47....1S,1984ApJ...287..255N,
1996AJ....112.1511S, 2002PASP..114.1097S, 
2003PASP..115.1211S, 2008AJ....136.2522M} as well as reduction in
\el{12}{C}/\el{13}{C}  
\citep{2003AJ....125..794P,2003MNRAS.345..311P}  along the RGB. Here we extend
this work and compare M3 to its second parameter partner M13 thus
testing the mechanism in another cluster of similar metallicity. We also turn to
the more extreme end of the GC metallicity distribution by investigating the
clusters M92, M15 and NGC 5466 (all with [Fe/H]$\approx-2.2$). This sample
allows
us to
investigate the mechanism in multiple clusters at a range of metallicites. 
The exercise is somewhat complicated by the fact that globular clusters possess
multiple stellar populations. 

Historically globular clusters (GCs) have been utilized as test-beds for stellar
theory. Some of the earliest published studies of GCs 
\citep{1952AJ.....57....4A, 1955AJ.....60..317A, 1953AJ.....58...61S}  coincided
with the first stellar evolution calculations
\citep{1952ApJ...116..317O,1962ApJ...135..770I}. The initial mass spread of the
stars and their presumed coeval nature contributed to the color magnitude
diagram's  usefulness as a diagnostic tool for the developing theory
\citep{1954MSRSL...1..254S,1955ApJ...121..616J,1968ApJ...153..101I}. During this
time GCs were considered to be simple stellar populations; that is, the
stars in
any given cluster were assumed to be of the same age and composition. Hence the
cluster color magnitude diagram was an
outcome of variations in stellar mass. The first study to challenge the
simple
stellar population
hypothesis was that of \citet{1947ApJ...105..204P}. He measured CN
band strength in the stars of M3 and M13 and found a CN-strong star amongst
many CN-weak. 
The importance of Popper's find was highlighted by \citet{1971Obs....91..223O}
who
found CN-strong stars in M5 and M10. 
These were the first clues that GCs possess
heterogeneous
C and N abundances. 

The heterogeneity of stars within GCs extends beyond carbon and nitrogen.
Along the RGB, as well
as on the main sequence, O, Na, Mg, and Al can display star-to-star variations
(e.g., \citealt{2004ARA&A..42..385G} and  \citealt{1994PASP..106..553K}
and references therein).
 It is now commonly
argued that  a previous generation of
stars polluted
the
medium from which a distinct second population formed. The first generation
imprinted the abundance patterns from the various stellar burning sites, whether
they were asymptotic giant branch stars \citep{2004MNRAS.353..789F},
super asymptotic giant branch stars \citep{2011MNRAS.410.2760V}, massive
binaries \citep
{2009A&A...507L...1D} or massive rotating stars
\citep{1993AJ....106..133B, 2006PASP..118.1225S, 2007A&A...464.1029D,
2010IAUS..266..131C}. This scenario is consistent with the
presence of C-N-O-Na abundance dispersions on the main sequences of GCs.

The formation of a second generation of stars from gas that is rich in
CN-processed material provides a paradigm that explains the observations of 
dichotomous CN band strengths
in GC stars. The CN band strength is a useful
indicator of nitrogen content of globular cluster RGB star atmospheres.
New stars that form from the polluted gas are inherently enriched in nitrogen as
well as \el{4}{He} compared to the primordial generation
\citep{1981ApJ...244..205N}. The fact that the dichotomy in the CN band strength
has been
detected below the bump in the luminosity function (LF bump), albeit for
clusters that are of intermediate to high metallicity, provides strong evidence
for the pollution scenario
\citep{1980ApJ...238L.149H, 1991ApJ...373..482B, 1991ApJ...381..160S,
1994A&A...290...69B, 1998MNRAS.298..601C,
1999AJ....117.2428C,2001AJ....122..242B, 2010A&A...524A..44P}.

The CN bands can be used as population tracers in all but the most metal-poor
clusters where [Fe/H] $< -2$ (e.g., \citealt{2010AJ....140.1119S,
2011AJ....142..126S}), since the CN bands are weak at low metallicity
\citep{2005AJ....130.1177C}.
In such cases the existence of multiple populations can be inferred through
other means. \citet{2000AJ....120.1351S} and
\citet{1997AJ....114.1964S} have
shown that in metal poor clusters [Na/Fe] varies by over 1 dex and this
variation is not correlated with RGB evolution. The temperatures required
for processing of Na are beyond those reached by thermohaline mixing on the RGB.
We therefore
expect a normal (primordial) and enriched population given the large spread.
This is a
strong indication that multiple populations do exist in these clusters.

Evidence of this multiple population scenario is not limited to the observations
of CN band strength or variations in [Na/Fe]. It is discussed at length by
\citet{2002AJ....124..364C},
\citet{2009ApJ...697L..58A}  and \citet[see
references therein]{2009IAUS..258..233P} 
for the main sequence (MS) and sub giant branch. Techniques such as
isochrone fitting require distinct \el{4}{He} abundances in these populations to
explain color magnitude diagrams. Examples of such clusters include: Omega
Centauri
\citep{2005ApJ...621..777P,
2005ApJ...621L..57L, 2007ApJ...654..915S}, NGC 2808
\citep{2005ApJ...631..868D,2005ApJ...621L..57L,2007ApJ...661L..53P} and 47
Tucanae \citep{2009ApJ...697L..58A}. We note also that calculations by
\citet{2008MNRAS.390..693D} show the horizontal branches of various clusters can
be best reproduced with the presence of two populations, one of which has 
elevated \el{4}{He} abundances, which is a prediction of the pollution
scenario.                                                                     

The prevailing picture is that in order to match observations, pollution from a
primordial generation must be present in the cluster to provide an enriched
environment from which a second population forms. Furthermore extra mixing must
be
operating in each generation \citep{1991ApJ...381..160S, 1998A&A...333..926D,
1999AAS...195.0303B,2002PASP..114.1097S} as they ascend the RGB. This is exactly
what
\citet{2002PASP..114.1097S} suggested is happening in M3.
\citet{2011ApJ...728...79A}
provided further evidence for this hypothesis and  inferred that the cluster is
comprised of two populations\footnote{This is not a bold inference. It is
expected that almost all globular clusters contain
multiple
populations. However this fact has not been \textit{observationally} confirmed
below the
the RGB of M3.} with different ages
and abundances. It was found that thermohaline mixing can account for the
variation in carbon
abundance of both the CN-weak and CN-strong stars but only for the nitrogen
abundance of the CN-weak stars. The CN-strong stars  have so much nitrogen to
begin
with that any extra mixing does not significantly affect the surface
composition.
Furthermore the spread in [N/Fe] is dominated by the variation that was present
among the cluster stars before they commenced RGB evolution.

In this work we examine the hybrid picture for the chemical evolution of GC
giants across a range of metallicity. We investigate the variation of surface [C/Fe] with magnitude on the giant branch of various clusters. The initial abundances are chosen to match the observed subgiant values. We concentrate on C and N because only for these species do data exist over a wide range of luminosity for clusters of different metallicity (data for Li as a function of magnitude exists for M4 and NGC 6397, but the lithium values saturate over a range $< 1$ magnitude whereas the [C/Fe] show variation over $ > 2$ magnitudes, see \citealt{2011MNRAS.412...81M} and \citealt{2009A&A...503..545L}. Also measurements of lithium above the LF bump are scarce and many are upper limits rather than detections.). In the intermediate-metallicity clusters
we compare M3 to M13. We also turn to the metal poor regime where
\citet{2010MmSAI..81.1057A} alluded to
significant problems with the
contribution of extra mixing in M92; namely the stars appear to begin mixing
before they reach the
LF bump.  Here we examine a further two similarly metal-poor
clusters in M15 and NGC 5466 in order to help identify
whether the behavior is due to a unique mixing history in M92 or is common
amongst
metal-poor systems.

\section{Calculations}
We use MONSTAR (the Monash version of
the Mt. Stromlo evolution code; see \citealt{2008A&A...490..769C}) to 
produce stellar models for the  clusters. Our implementation of thermohaline
mixing
follows that of \citet{2011ApJ...728...79A},
\citet{2010A&A...522A..10C}, \citet{2010MNRAS.403..505S},
\citet{2009MNRAS.396.2313S} 
 and 
\citet{2007A&A...467L..15C,2007A&A...476L..29C} in that we use the formulation
developed by \citet{1972ApJ...172..165U} and 
\cite*{1980A&A....91..175K}, where
thermohaline
mixing is modelled as a diffusive process with coefficient:

\begin{equation} \label{eq:kip}
D_t =  C_t \,  K  \left({\varphi \over \delta}\right){- \nabla_\mu \over
(\nabla_{\rm ad} - \nabla)} \quad \hbox{for} \;  \nabla_\mu < 0 ,
\label{dt}
\end{equation}

where $\varphi = (\partial \ln \rho / \partial \ln \mu)_{P,T}$, 
$\delta=-(\partial \ln \rho / \partial \ln T)_{P,\mu}$, 
$\nabla_{\rm{\mu}} = (\partial \ln \mu / \partial \ln P)$, 
$\nabla_{\rm{ad}}=(\partial \ln T / \partial \ln P)_{\rm{ad}}$, 
$\nabla = (\partial \ln T / \partial \ln P)$, 
\textit{K} is the thermal diffusivity and $C_t$ is a dimensionless free
parameter.
In fact $C_t$ is related to the aspect ratio, $\alpha$, of the 
thermohaline fingers (assumed to be cylindrical) by:

\begin{equation} \label{eq:ct}
C_t = {8 \over 3} \pi^2 \alpha^2.
\end{equation}

Following \citet{2011ApJ...728...79A}, we employ the empirically
derived value of
\textit{C$_{t}$}$= 1000$ for the stars in GCs. It should be noted 
that \citet{2010A&A...521A...9C} and \citet{2010ApJ...723..563D} prefer a lower
value of \textit{C$_{t}$}$= 12$. Their choice of 
parameter is based on theoretical grounds and matches the choice of
\citet{1980A&A....91..175K}. Numerical simulations from    \citet{2011ApJ...727L...8D} and \citet{2011ApJ...728L..29T}  also indicate that a value of \textit{C$_{t}$}$= 12$ should be adopted.
This value is unable to reproduce observations \citep{2007A&A...467L..15C, 2011ApJ...728...79A} but must serve as a caveat to the approach taken in this work. We return to this point in Section \ref{rolethm}.

\subsection{Intermediate metallicity: M3 and M13}
\citet{2011ApJ...728...79A} showed that the chemical evolution of the stars in
M3 is consistent with the hybrid theory of globular cluster evolution. In
intermediate metallicity clusters the contribution of primordial enrichment can
be traced through measurements of
the CN bands and these were used in M3 to infer the
distinct populations. Furthermore in each population the evolution of [C/Fe]
along the RGB was well modelled by
thermohaline mixing. It is important to test whether this result
    pertains to other clusters of similar metallicity. As such we turn our
attention to M13.
M3 and M13 are a second parameter pair of metallicity [Fe/H]$\approx	 -1.4$
and ages
between 11.3 and 14.2 Gyr
\citep{1992ApJ...394..515C,1996MNRAS.282..926J,2000ApJS..129..315V,
2002A&A...388..492S,2004AAS...205.5301A}. Similarities between the two clusters
exist in age, metallicity, Na-O anticorrelation 
\citep{1992AJ....104.2121S, 1992AJ....104..645K}, Na spread 
\citep{1978ApJ...223..487C, 1980ApJ...237L..87P, 1992AJ....104..645K} and
r-process variation \citep{2011ApJ...732L..17R}; it is only horizontal branch
morphology that differs significantly. We therefore expect
thermohaline mixing operating in multiple populations to also model the
variation of [C/Fe] (and [N/Fe]) in M13.  

We use data for M3 and M13 from a variety of sources. The M3 data are taken
from \citet{2002PASP..114.1097S} which is compiled from previous studies in the
literature, namely \citet{1981ApJS...47....1S}, \citet{1996AJ....112.1511S} and
\citet{1999AJ....118..920L} with a zero point offset applied in order to
homogenize the data. In addition to determining  carbon and nitrogen abundances
as a
function of absolute magnitude, \citet{1981ApJS...47....1S} and
\citet{1984ApJ...287..255N} measured the CN strength for M3 giants.
We do not possess measurements
of CN strength for the stars in the M13
sample but (unlike M3) observations of [C/Fe] and [N/Fe] exist on
the main sequence. For M13 we incorporate the studies of \citet{1981ApJS...47....1S},
\citet{1996AJ....112.1511S}, \citet{2002ApJ...579L..17B},
\citet{2004AJ....127.1579B} and \citet{2005AJ....129.1589S}. Offsets to allow
for systematic differences have been
applied by \citet{2004AJ....127.1579B} for all but the most recent study.
We assumed a distance modulus of $m-M=14.43$ which is an average of various measurements
($m-M=14.44$ from  \citealt{1992MNRAS.257..731B}, $m-M=14.33$ from
\citealt{1996AJ....112.1487H}, $m-M=14.47$ from \citealt{1997ApJ...491..749G},
$m-M=14.48$ from \citealt{1997AJ....114..161R}). The uncertainty introduced here is no
larger than the
measurement error of the observations. 

\begin{figure*}
\begin{center}
 \includegraphics[scale=0.3]{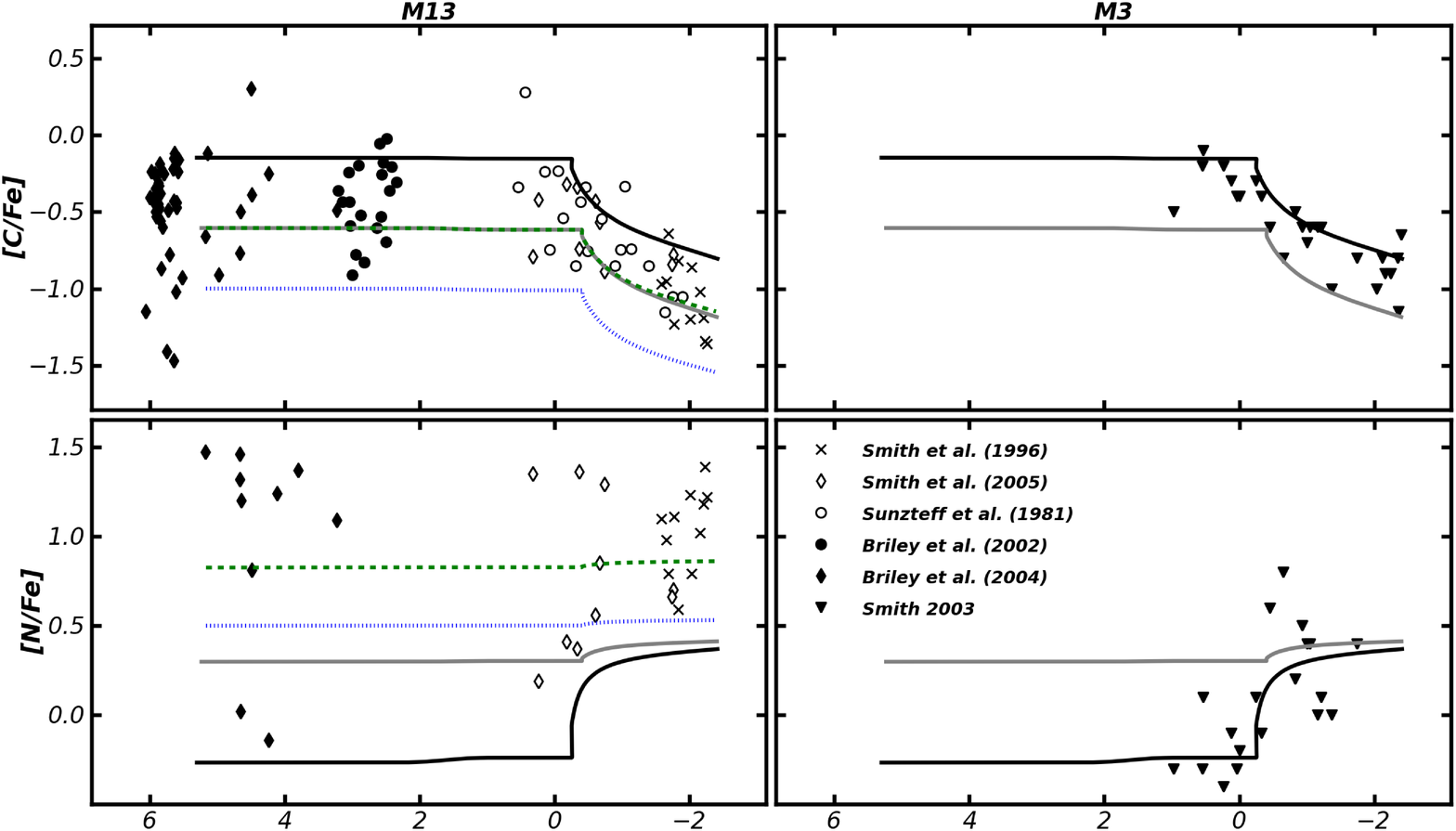}

\textbf{M\boldmath$_V$}                        \end{center}
\caption{[C/Fe] and [N/Fe] vs $M_V$ for M13 (left panels) and M3 (right panels).
The various studies from  which the data are taken are listed in the Figure. In
all panels the solid  black curve represents our CN-weak model and the solid
grey curve
corresponds to our CN-strong model . For M13 we also
provide a nitrogen enhanced model (green dashed line) and a model designed to
match the main-sequence carbon spread of the cluster (blue
dotted line). In all models the metallicity
is set to $Z = 0.0005$ and the
thermohaline
mixing parameter \textit{C$_{t}$}$ = 1000$.}
 \label{fig:m3m13}
\end{figure*}

In Figure \ref{fig:m3m13} we  plot the determined [C/Fe] and [N/Fe]
values
against absolute visual magnitude for M13 and M3. Symbols are
used to specify to which study the data corresponds; the legend can be
found in the figure. We
include two
calculations (the
two solid curves) used to model M3 (see \citealt{2011ApJ...728...79A}). The
models are of mass $M = 0.8 \Mo$ and metallicity $Z = 0.0005$. In these
calculations we run without convective overshoot, our
mixing
length parameter $\alpha$ is set to 1.75 and thermohaline mixing parameter
\textit{C$_{t}$}$= 1000$. 
In both clusters the solid  black curve represents our
CN-weak model with initial abundances
$Y = 0.2495$, $X(C) = 5.45 \times$10$^{-5}$,  $X(N) = 1.5 \times$10$^{-5}$ and
$X(O) = 2.86 \times$10$^{-4}$. The initial carbon and nitrogen values were
selected to match the measurements of these species in the CN-weak stars in M3. 
The
solid grey curve corresponds to our CN-strong
model with $X(C) = 1.9 \times$10$^{-5}$, $X(N) = 5.5 \times$10$^{-5}$ and
$X(O) = 2.6 \times$10$^{-4}$ to match the CN-strong population in M3. As the
stars are likely to have undergone CN cycling
we
increased our \el{4}{He} to $Y = 0.28$ which only has a marginal effect on the
location of the LF bump.

Thermohaline mixing operating in two populations models
the behavior of [C/Fe] along the RGB in both M3 and M13. In the case of M13 we
have the added constraint of observations along the main-sequence.
In this cluster the CN strong model acts as a lower envelope to the RGB
observations and
simultaneously accounts for most of the main-sequence data. 
However, there are a number of sub-giant and main-sequence stars 
at lower [C/Fe] than seen in the M3 data.  In order to match the entire
main-sequence
spread we have taken our CN strong model and  reduced the initial carbon by a
further 0.4 dex. This is our dotted blue curve where we set initial abundances
to $X(C) = 7.7 \times$10$^{-6}$, $X(N) = 8.8 \times$10$^{-5}$ and
$X(O) = 2.4 \times$10$^{-4}$. By design this model matches the most carbon
depleted subgiants (except for
the three extreme cases with [C/Fe] $< -1$) but the predicted abundances along the RGB
are systematically lower than the current observations. It appears as if these most-carbon-depleted subgiants lack counterparts on  the RGB. This could be
the role of small statistics or the effect of systematic offsets from
different data sources; recall we do not apply an offset for the
\citet{2005AJ....129.1589S} study. We are unable to account for the 
two most carbon-poor stars in M13 ([C/Fe] $\approx -1.5$). These stars are at the faint
end of the \citet{2004AJ....127.1579B}  catalogue and hence have large associated errors (Cohen 2011, private communication).

Compared to M3, there appears to be
a larger star-to-star variation in nitrogen in M13; a significantly N-enhanced
model is required to reproduce the
upper envelope of the data.  In the context of
the current paradigm the enrichment is proposed to be due to ON cycling in
the polluting generation of stars. To investigate this we constructed an
ON-cycled model (dashed green line) with half the oxygen converted to N, giving
$X(C) = 1.9 \times$10$^{-5}$, $X(N) = 1.85 \times$10$^{-4}$ and
$X(O) = 1.3 \times$10$^{-4}$. The figure shows that the change in [N/Fe]
produced by thermohaline mixing is negligible. The initial nitrogen in these
stars is sufficiently high that
any nitrogen brought to the surface via thermohaline mixing makes little
difference to the envelope composition.

We note from Figure \ref{fig:m3m13} that even the ON cycled model (green dashed
line)
fails to account for the N-rich stars which show 1 dex more N than this model.
Even processing all the O into N would not produce a sufficiently high N
abundance. We must consider the possibility, therefore that C+N+O is higher in
M13 than in M3. This is a testable prediction. 

As we have stated there are a
larger number of stars in M13 that are  enriched in nitrogen;
\citet{2011AJ....142..126S} find the cluster
contains four times as many CN-strong stars than CN-weak ones.
In terms of the hybrid theory this is interpreted as the cluster containing a
majority of second generation stars. In \citet{2002PASP..114.1097S} and
\citet{2011AJ....142..126S} the observations suggest that the M3 stars are
evenly split amongst the CN-weak and CN-strong populations. 
The inferred ratios of the stellar populations within the clusters,  determined
from measurements of the CN bands agree with the theoretical predictions of     
\citet{2008MNRAS.390..693D}. They have independently determined that
more than 70\% of stars in M13 and 50\% in M3 are required to be enriched in
\el{4}{He} (a product of CN cycling along with N enrichment) to reproduce
the respective horizontal branch morphologies.

In M3 there is a distinct lack of CN-strong stars at low luminosity. As
discussed in \citet{2011ApJ...728...79A} 
this is an artifact of the original \citet{1981ApJS...47....1S} study in which
the lower luminosity stars that were observed happened to be CN-weak.
\citet{1984ApJ...287..255N} showed that CN-strong giants do exist in M3 at
luminosities corresponding to the faint limit of the \citet{1981ApJS...47....1S}
survey, as did \citet{2011AJ....142..126S}. The latter show the CN band strength
remains clearly dichotomous in M3 even at low luminosity. We assume this will 
be reflected in the  
[N/Fe] abundances and therefore [C/Fe]. So whilst we possess measurements of the
CN bands
for our M3 sample and a few observations of [C/Fe] at low luminosity,
conversely in  M13 we possess many low-luminosity observations of [C/Fe] but
no
measurements of the CN bands. Once again the work of
\citet{2011AJ....142..126S} provides some insight. They demonstrate an obvious
dichotomy in the CN bands along the RGB of M13. The dichotomy is not as clear
below the
sub
giant branch as the temperatures are too high to allow molecule formation.
\citet{2011AJ....142..126S} also show a greater spread in the CN band values
especially in the CN strong stars; this is reflected in the large spread in
[N/Fe] observed in the cluster. 

\subsection{Low Metallicity}
\subsubsection{NGC 5466}
\begin{figure*} 
\begin{center}
 \includegraphics[scale=0.3]{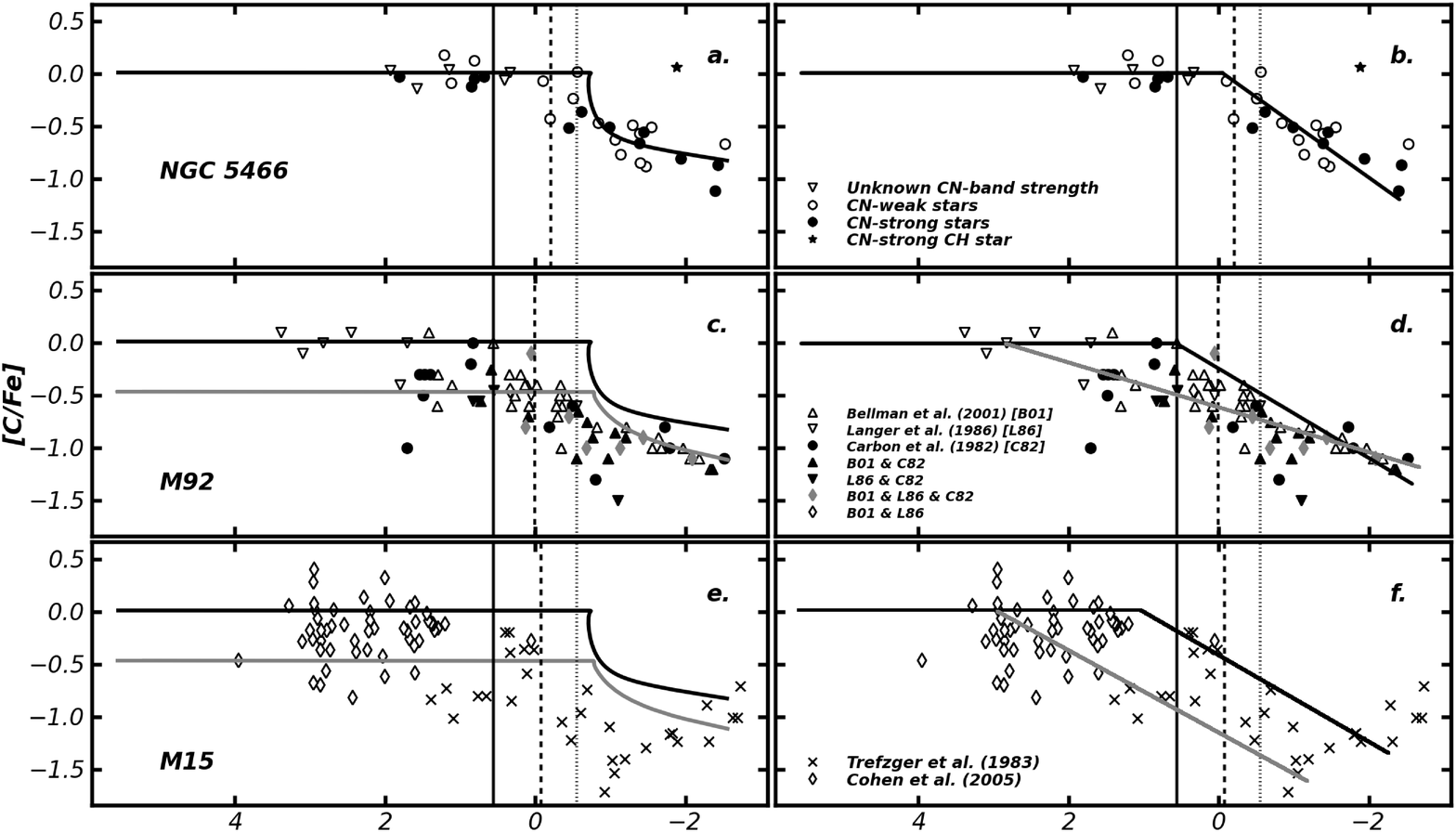}

 \textbf{M\boldmath$_V$}
\end{center} 
\caption{[C/Fe] vs $M_V$ for NGC 5466 (top panels), M92 (middle panels) and
M15 (bottom panels). In the left panels the black curve represents our solar
scaled model (CN-weak) and the grey curve is our CN-strong model for these
clusters.
The
solid vertical line represents the end of first dredge-up according to the
models. The
dotted line corresponds to the LF bump according to \citet{2008AJ....136.2522M}.
The dashed lines are photometric derivations of the LF bump for each cluster:
for M92 and M15 the magnitude is taken from \citet{2011arXiv1109.2118N},
 and for NGC 5466 from \citet{2007ApJ...663..277F}.
In the right panels we provide lines of best fit by eye which helps highlight
the strange behavior of the clusters.}
 \label{fig:m92}
\end{figure*}

Carbon depletion along the RGB of
NGC 5466 has been observed by \citet{1985A&A...145...97B},
\citet{2007ApJ...663..277F} and \citet{2010AJ....140.1119S}. In this study we
use the data from  \citet{2010AJ....140.1119S} as the observations are from a
single instrument, they cover a large luminosity range and for each star 
[C/Fe] and the CN band strength were determined. As was the case for the
intermediate-metallicity clusters, we plot [C/Fe] as a function of absolute visual
magnitude. In  Figures \ref{fig:m92}a and  \ref{fig:m92}b the NGC 5466 data are
plotted according to CN band strength; open
circles denote CN-weak stars, filled circles denote CN-strong stars\footnote{The
solid star is CN strong but it is most likely a CH star and hence a
binary.} whilst triangles represent stars where the CN band
strength is unknown. The
black solid curve in Figure \ref{fig:m92}a is our thermohaline mixing model with
initial
abundances of $Y = 0.25$, $X(C) = 1.6 \times 10^{-5}$,
$X(N) = 5.03 \times 10^{-6}$ and $X(O) = 4.57 \times 10^{-5}$.  The model is of
mass $M = 0.8 \Mo$ and
metallicity $Z = 0.0001$ corresponding to the observed [Fe/H]$= -2.2$ and age
between 11 and 16 Gyr \citep{2010PASP..122..991D, 2000AJ....120.1884G}. We run
without convective overshoot, set the
mixing
length parameter to $\alpha = 1.75$ and set the thermohaline mixing parameter
to \textit{C$_{t}$}$= 1000$. Highlighted also are the locations of the important
mixing events along the RGB.
Common to all panels are the solid vertical line and dotted line. 
The
solid vertical
line  represents the end
of first dredge-up as calculated in the models; this occurs at $M_V \approx
0.56$.   The
dotted line corresponds to the LF bump at [Fe/H]$ = -2.2$ according to
\citet{2008AJ....136.2522M}
based on a metallicity-LF bump relation; this occurs at $M_V \approx -0.55$.
For each individual cluster we also include the dashed line which represents a
photometrically derived value of the LF bump magnitude. In NGC 5466
this was determined by \citet{2007ApJ...663..277F} and occurs at $M_V \approx
-0.2$. The end of first dredge-up marks the end 
of any surface composition changes expected by canonical theory. For stars of this
 mass and metallicity there is no 
visible change to the envelope composition from first dredge-up: the change in
surface mass-fraction of carbon is $ 6\times 10^{-8}$ in our models.  The
position of the LF bump
represents the earliest point at which thermohaline 
mixing can begin to operate. It corresponds to the point when the hydrogen
burning shell meets the composition discontinuity left 
behind by first dredge-up.  

Figure \ref{fig:m92}a suggests that our solar scaled
model (black curve) is a good fit to observations of [C/Fe] in NGC 5466. 
Surface depletion in 
the cluster appears to begin after the  LF bumps determined by
\citet{1990A&A...238...95F} and \citet{2008AJ....136.2522M}.
We find that the bump determined by \citet{2008AJ....136.2522M} is a more
convincing fit to the models than that of
\citet{2007ApJ...663..277F}. \citet{2007ApJ...663..277F} also inferred
the LF bump magnitude by isochrone fitting and obtained a value 0.3 magnitudes
brighter agreeing with the \citet{2008AJ....136.2522M} prediction.
The determination of the LF bump at low
metallicity is difficult. We return to this point in Section
\ref{compz}

Although a single CN population undergoing thermohaline mixing is a
satisfactory fit to the data, in Figure \ref{fig:m92}b
we
provide an equally plausible fit drawn by eye. The
black solid line (which applies to Figures \ref{fig:m92}b, \ref{fig:m92}d \&
\ref{fig:m92}f) tries to take into account our
belief that carbon depletion
should not occur until after first dredge-up but does not take into
consideration any preconceptions of the mixing profile. 
If we assume the bump luminosity calculated by \citet{2007ApJ...663..277F} is
correct then depletion begins after the LF bump, as expected. If, however, we
use the bump luminosity calculated by \citet{2008AJ....136.2522M} then the
mixing may be argued to begin before the bump is reached.

NGC 5466 requires almost no spread in the initial composition to account for the
spread of
carbon. In addition
to measurements of [C/Fe], \citet{2010AJ....140.1119S} analyzed the CN bands of
the stars in their
sample. The results hinted that a CN dichotomy, which is typical in globular
cluster red giants, may be present but as we can see from Figure \ref{fig:m92}a
the stars do not separate as clearly as those in the more metal rich clusters.
\citet{2010AJ....140.1119S} used a linear least-squares fit in the plane of the
S(3839) CN index \citep{1981ApJ...244..205N} versus absolute V magnitude to
divide relatively CN-strong stars (those above the line) from CN-weak stars
(those below the line). As in other studies of low-metallicity globular clusters
(e.g., \citealt{2008PASP..120....7M}), the mean separation in CN band strength
between the two groups is not large relative to the scatter within each group.
Whilst the data suggest that there is a CN bimodality in NGC 5466, its low
metallicity reduces the effectiveness of CN band strength as a marker of
multiple populations in this cluster.

\subsubsection{M92}

The M92 data in Figures \ref{fig:m92}c and \ref{fig:m92}d are those adopted by 
\citet{2003PASP..115.1211S} and comprise data from various sources to which
offsets have
been
applied in order to remove systematic
differences in abundance scales. These original sources are
studies by 
\citet{1982ApJS...49..207C}, \citet{1986PASP...98..473L} and
\citet{2001PASP..113..326B} and we highlight the target stars in each catalogue (see the legend in the figure). Even with typical errors of $\pm 0.2$ dex there is a much larger spread in [C/Fe] at any given magnitude than was seen in NGC 5466. In these M92 panels the dotted line is the
photometrically derived LF bump identified by \citet{2011arXiv1109.2118N}. These authors have undertaken high resolution HST observations to determine the LF bump in many globular clusters. Their measured value of the magnitude of the bump $M_V \approx 0.016$. ($V_{bump} = 14.666 \pm 0.013, \ \ (m-M)V=14.65$) differs from the older \citet{1990A&A...238...95F} value of  $M_V \approx -0.39$. We do not possess measurements of CN band
strength
for the M92 stars plotted in   
Figures \ref{fig:m92}c and \ref{fig:m92}d but \citet{1985ApJ...299..295N} and
\citet{2011AJ....142..126S} have provided evidence that a bimodality may exist.
As
is the case for NGC 5466 the separation in the populations is marginal, a
property that appears common at low metallicity. Even without the CN bands
the spread in [Na/Fe] \citep{2000AJ....120.1351S, 1997AJ....114.1964S}, and the
spread in [C/Fe] at low luminosities is evidence for mixing between the two
populations.

The depletion of [C/Fe] in M92 has previously been
investigated  by \citet{2003ApJ...593..509D} but over a smaller luminosity
range. Their (canonical) extra mixing formulation requires two models, one
depleted by 0.1 dex and the other by 0.5
dex in [C/Fe] to match the abundance spread. \citet{2010MmSAI..81.1057A}
considered the depletion of [C/Fe]
with a
different implementation of thermohaline mixing to that used here. They
highlighted that M92 
appeared to show surface depletion of carbon before FDU, a property 
in disagreement with not just thermohaline mixing but also canonical stellar
theory.
This strange behavior
has been
discussed by (at least) \citet{2008AJ....136.2522M},
\citet{2004ARA&A..42..385G}
\citet{2001PASP..113..326B}, and
\citet{1986PASP...98..473L}.

This apparent pre-FDU depletion is highlighted in Figure
\ref{fig:m92}d. Here the
black solid line is an  eye fit based on the assumption that carbon depletion
should not occur until after FDU while the grey curve is a best fit
by eye to the data without any preconceived luminosity for the onset of carbon
depletion. If we follow the latter fit, then by the time the end of FDU is
reached the
stars have
depleted almost 0.5 dex in [C/Fe]. This is unusual because, as we have stated,
at
this mass and metallicity we expect changes due to FDU to be of the order 
$\Delta \rm{X_{Surface}}$(C) $= 6\times
10^{-8}$. Between the end of first dredge-up ($M_{\rm{V}}$ $\approx +0.5$) and
the determined location
of
the LF bump according to \citet{2008AJ....136.2522M} ($M_{\rm{V}}$ $\approx -0.5$; the earliest point at which
thermohaline mixing is expected to begin) the stars 
in M92 are depleted by roughly a further 0.3 dex in [C/Fe]. Depletion is present but not as pronounced if the \citet{2011arXiv1109.2118N} bump magnitude is used.  It is
possible that
these are the observable effects of other 
extra-mixing mechanisms and this may also explain the
carbon depletion 
prior to the end of first dredge-up. If true, this would have significant
implications
for stellar
theory. First of all such depletion was not seen 
in other clusters \citep{2003PASP..115.1211S, 2010AJ....140.1119S}.
In most scenarios, extra mixing is
inhibited until the star reaches the LF bump
and the advancing H-shell removes the molecular weight
discontinuity left behind by the receding convective envelope. In the case of
M92 some 
stars on the giant branch have already depleted their [C/Fe] by about 0.8
dex when the models reach this stage. If we have to postulate that some form of
mixing begins sufficiently early to produce this depletion, then the
mixing
must necessarily remove the abundance discontinuity
that is itself responsible for the LF bump. If the apparent behavior is real,
we believe this is a
significant problem for
stellar astrophysics. 

If we assume that M92 displays an
initial spread of [C/Fe]=0.5 dex then can we remove the issue of mixing
before the end of FDU?
The black curve in Figure \ref{fig:m92}c is the same
solar scaled model
applied to match the stars in NGC 5466. As we now require 
a greater spread in
carbon to match this cluster we include a second model (grey solid curve) run
with the same physical parameters but a composition that is reduced in C and enhanced
in N. For this model we set $Y = 0.25$,  $X(C) = 5.46 \times
10^{-6}$, $X(N) = 1.60 \times 10^{-5}$ and $X(O) = 4.57 
\times 10^{-5}$ giving [C/Fe]=-0.5. We have not changed Y in the model. 
The main effect of changing Y is to alter the the location of the LF bump and as
we have seen in
M3 and M13 this is negligible. These two thermohaline models with large initial
carbon spread remove
the apparent pre FDU depletion. Assuming such a spread in [C/Fe]
raises the issue that there are no
stars observed with [C/Fe]=-0.5 below a magnitude of
$M_{\rm{V}} \approx 2$.  Only four stars have been observed at 
such magnitudes and all have [C/Fe] $\approx 0$. A large spread in [C/Fe] is
observed in M15 and M13 below the sub giant branch. We expect this is the case
in M92 but require a targeted study to confirm this assumption.

A large spread in [C/Fe] does not solve all the problems associated with M92.
Mixing still appears to be occurring before the LF bump,  even for the model
with an initial [C/Fe]=-0.5. There is approximately 1
magnitude between the location where the carbon abundances turn
down and where the models suggest depletion should begin. Of greater concern is
the inability of the models to match the upper RGB abundances of stars in
M92. The initial abundances were selected to cover the spread in [C/Fe] before
the onset of extra mixing. No stars that have depleted [C/Fe] (magnitude
brighter than $M_{\rm{V}} = 0$) fall along the evolutionary path predicted by
the solar scaled model (black curve). If such stars do exist and are
governed by thermohaline mixing, we expect them to be bright
members in the cluster (ascending the RGB). 
It would be perverse if by chance they were not selected for any of the
three high resolution
spectrographic studies (unless of course, they are all in the centre of the
cluster). The initially depleted model (grey curve) provides an upper envelope
to the mixed RGB data, however the composition was selected
to form a lower limit of the initial M92 abundance spread. It is sobering to
see how many data points fall below this curve. Although the two models with an
 initial spread of 0.5 dex in [C/Fe] can encompass fainter stars ($M_{\rm{V}} >
0.5$ there are very few stars between the curves once carbon depletion has
begun. Thermohaline does not deplete carbon to the levels seen in M92: the
models are unable to simultaneously account for both the sub-giant branch and
RGB data.  

The M92 sample is a combination of three different studies and although
attempts have been made to homogenize the data, systematic differences in the
studies still may lead to offsets of $\simeq 0.3$ dex in the [C/Fe] abundances.
At such
low metallicity one has to wonder if inhomogeneity in the data may be affecting
our interpretation of the mixing history. Extensive measurements of [C/Fe] at
various magnitudes in this cluster would provide a robust test for stellar
evolution models.

\subsubsection{M15}
[C/Fe] as a function of magnitude is plotted for M15 in Figures
\ref{fig:m92}e and \ref{fig:m92}f. The data are a combination of
those from \citet[crosses]{1983ApJ...266..144T} and
\citet[diamonds]{2005AJ....130.1177C}
and plotted  as given in these sources. No offsets have been applied to
correct for systematic errors so the data are marked according
to their source. As is the case with M92 we do not possess CN band
strength
measurements for our sample but  \citet{2011AJ....142..126S} show
there would be little separation between the mean values of the CN-strong and
CN-weak stars, a common aspect of the three
metal poor clusters studied here. The same is also true for (at least)
NGC 5053 a cluster with metallicity [Fe/H] $= -2.3$ 
\citep{2011AJ....142..126S}. In Figures \ref{fig:m92}e and \ref{fig:m92}f
the dotted vertical line represents the magnitude of the M15
LF bump determined by \citet{2011arXiv1109.2118N} which occurs at $M_V \approx -0.075$ ($V_{bump} = 15.315 \pm 0.021, \ \ (m-M)V=15.39$). This is in comparison to \citet{1999ApJ...518L..49Z} who also used HST data to determine a bump magnitude of $M_V \approx -0.42$.  

In Figure \ref{fig:m92}e we plot the two thermohaline mixing models
applied in M92. In both clusters thermohaline mixing
does not deplete carbon to the degree seen in the observations. Like M92
there is evidence that M15 is mixing between the end of FDU and the LF bump.
This is
highlighted by Figure \ref{fig:m92}f where we
plot our lines drawn by eye rather than the stellar models. This effect would be exacerbated had we adopted the location of the LF bump according to \citet{1999ApJ...518L..49Z}. It is possible this
is
an artefact of combining inhomogeneous data. There is some luminosity overlap
between the two studies and it is only when they are combined that it appears
as if pre bump depletion is occurring. This same argument can be used to explain
any pre-FDU mixing alluded to by the grey line of best fit in Figure
\ref{fig:m92}c. A homogeneous set of
observations over this luminosity range (as exists for NGC 5466)  is
required.

While similarities exist there are also slight differences between M15 and
M92.
A large spread of [C/Fe] is present in M15, which extends below the sub giant
branch.
This is justification for the inclusion of our initial [C/Fe] $= -0.5$ model
(grey curve in Figures \ref{fig:m92}c and \ref{fig:m92}e). In M92 it is the
dearth of observations at low luminosity that raises
many questions about the
behavior of the cluster and its initial abundance spread. We note there seems
to
be a lack of stars
with high [C/Fe] just before FDU in M15. We assume these stars exist and the
current void is a result of combining two studies that focus
on stars at different stages of evolution. Perhaps the most curious aspect of
M15 is
the behavior of the stars at $M_V \approx -1$. The stars appear to suddenly
increase their carbon abundance. We await
confirmation of this behavior before speculating on its cause as it is
inconsistent with our current understanding of stellar nucleosynthesis.  

\citet{2009A&A...503..545L} and
\citet{2011MNRAS.412...81M} use lithium to trace extra mixing along the
giant branches of NGC 6397 and M4 respectively. Such studies provide a
complimentary diagnostic to the variation of C and N with luminosity. 
Lithium is very sensitive to nuclear burning: it is destroyed at low
temperatures.
Any transport of material into warmer regions will be reflected through a
depletion in the surface lithium abundance.   
In the case of M15 and M92 this would hopefully help identify the magnitude at
which extra mixing begins in these clusters.
Given the behavior of M15 and M92 the point at which surface depletion
begins (after first dredge-up) must be considered distinct from the LF bump
(even though
there are good theoretical reasons why the two should coincide as they seem to
at
higher metallicity). For our
purposes we can only rely on high level photometry (and statistical analysis) of
the clusters to distinguish the location of the bump and, as we discuss in
section \ref{compz}, this
can be difficult.

\section{Discussion}

\subsection{Comparison across metallicity} \label{compz}
In our two intermediate-metallicity clusters the spread in [C/Fe] is explained
by
distinct populations which clearly separate according to CN band strength. This
is not
the case in low-metallicity clusters \citep{2010AJ....140.1119S,
2011AJ....142..126S}. Any CN bimodality is
marginal at best. The difference between the mean value of a CN index such as
S(3839)
between CN-strong and CN-weak stars is so small
that their distribution can be interpreted as a single CN population
with scatter. There is still clear evidence for multiple populations in the
metal poor clusters from the large spreads in O, Na, Al abundance within them.
We therefore expect that a polluting generation will enrich
the cluster with CN processed material and \el{4}{He}. This should be reflected
in the carbon and nitrogen abundances. In M15 there is a large spread in both
[C/Fe] ($\approx$ 1.5
dex) and
[N/Fe]  \citep[$\approx$ 3 dex,][]{2005AJ....130.1177C} but only a marginal
change in the CN
bands \citep{2011AJ....142..126S}. In these low metallicity clusters 
the CN band strength is not necessarily representative of the nitrogen abundance
in the stars.

According to our stellar models with thermohaline mixing, the clusters studied
here are close enough in metallicity that we should expect a similar degree of
carbon depletion.  In the intermediate metallicity
clusters (M3 and M13), we are able to match
the depletion of [C/Fe] along the RGB using two models of different initial
abundances. In the metal poor systems thermohaline mixing can account only for
the evolution of [C/Fe] in NGC 5466. In M15 and M92 not only do the clusters
appear to have depleted far more carbon than we predict but this depletion
has begun before the LF bump. Such behavior is not only inconsistent with
thermohaline mixing but also standard stellar evolution. 

It is worth noting that
the LF bump in extremely metal-poor clusters (e.g., M92, M15 and NGC 5466) is
not
as clearly visible as it is in more metal-rich clusters. There are two
theoretical reasons why this is so. Firstly the depth of first dredge-up is a
function of metallicity. Metal-rich stars  have
deeper convective envelopes than metal poor stars and therefore a greater
discontinuity in the
hydrogen profile that develops after first dredge-up. The burning shell in
metal-poor stars will encounter a smaller hydrogen difference and spend less
time
readjusting the stellar structure. Because less time is spent at the magnitude
of the
bump  the likelihood
of observing stars at this location is reduced. Secondly, evolution on
the RGB
speeds up as stars move towards brighter luminosities, so the expected number of
stars scales inversely with the luminosity in a volume limited sample (but not
in a magnitude limited sample). Hence the higher the luminosity at which the
bump occurs (i.e., the lower the metallicity), the lower the overall number of
stars one should expect to observe and the harder it is to identify the bump.

\citet{1990A&A...238...95F} identified an LF bump in M92 by co-adding data for
three very similar clusters. Their
determined
value of $M_{\rm{V}}$ $\approx$ $-0.39$ differs slightly from that of
\citet{2008AJ....136.2522M} who find $M_{\rm{V}}$ $\approx$ $-0.55$ based on the
use of
multiple clusters to determine  a LF bump-metallicity relation. 
Work by
\citet{2007AJ....133.2787P} provides little
evidence for a bump in the observed LF of M92 however recent work by \citet{2011arXiv1109.2118N} using HST data suggests that it is present at a magnitude of $M_{\rm{V}}$ $\approx$ $0.016$. For NGC 5466
\citet{2007ApJ...663..277F} identified a bump using statistical arguments. The
bump in M15 was more easily identifiable through the use of HST data as was done
by
\citet{2011arXiv1109.2118N} and \citet{1999ApJ...518L..49Z}.

\begin{figure*}
\begin{center}
 \includegraphics[scale=0.3]{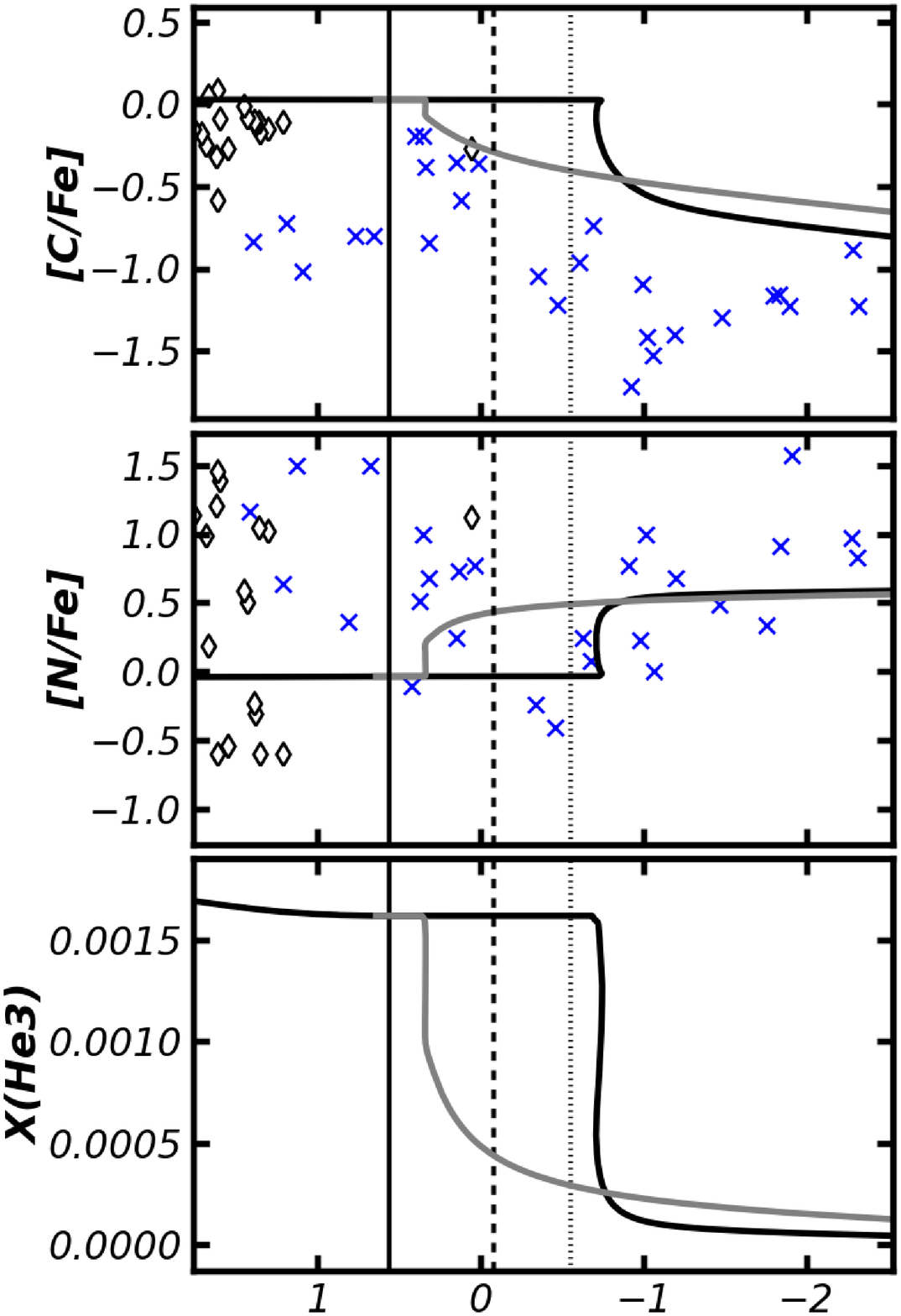}

\textbf{M\boldmath$_V$}                        
\end{center}
\caption{[C/Fe] [N/Fe] and X(\el{3}{He}) vs $M_V$ for M15. As above the
observational data is
taken from \citet{1983ApJ...266..144T} (crosses) and
\citet{2005AJ....130.1177C} (diamonds). The solid black curve corresponds to the
thermohaline mixing model with standard dredge-up depth (see Figure
\ref{fig:m92}) where the envelope extends down to a depth of $M = 0.368 \Mo$.
The
grey curve represents our model where we have artificially homogenized beyond
the depth of first dredge-up to a mass of $M = 0.320 \Mo$. The solid vertical
line represents
the end of first dredge-up according to the normal models. The dotted line
corresponds to the LF
bump of \citet{2008AJ....136.2522M} based on a metallicity-LF bump relation.
This occurs at $M_V \approx -0.55$. The dashed line is a photometric derivation
of the LF bump taken from \citet{1999ApJ...518L..49Z} and occurs at $M_V \approx
-0.42$.}
 \label{fig:deep}
\end{figure*}

One possible cause for altering the location of the bump (and hence the onset
of thermohaline mixing) in our models is to decrease the
\el{4}{He} abundance \citep{1978IAUS...80..333S}. However, extrapolating from
the \citet{1978IAUS...80..333S} results  to match the
observations would require an initial \el{4}{He} well below the Big
Bang
Nucleosynthesis value. Alternatively, if first dredge-up penetrates deeper than
the models predict,
then the hydrogen
burning shell will encounter the homogenized region at lower luminosities. In
Figure \ref{fig:deep} we plot our metal-poor model with solar scaled abundances,
in which the
convective envelope extended down to a mass of $M = 0.368$ $\Mo$. We have
included
a model (grey line) with the same initial abundance but artificially homogenized
(mixed from the surface) down to a
mass of $M = 0.320$ $\Mo$ at the time of FDU. The results show that extending
the depth of first
dredge-up by
about 15\% in mass translates to the mixing beginning about one magnitude
fainter. Note that the artificial model undergoes a period of adjustment
as it returns to the structure of the standard metal-poor model. 
During this brief phase the burning shell is
closer to the homogenized region than the structure would
dictate if deeper dredge-up occurred normally. As a consequence the
temperatures are such that \el{3}{He} is easily destroyed but carbon depletion
is still inefficient; this can be seen in Figure \ref{fig:deep}.
\cite{2011ApJ...728...79A} call this situation a 
``burning limited'' regime. The material is efficiently transported to the
desired regions but the processing is inefficient due to the burning
conditions. The artificial model therefore has less \el{3}{He} to drive the
mixing near the tip of the RGB when the conditions are more conducive to carbon
depletion, a regime \cite{2011ApJ...728...79A} call ``transport limited''. In
this scenario the inefficient transport of material to the burning regions is
the limiting factor in the processing. This however
is a side issue stemming from the artificial model; the important point is that
one way for the models to match
the
observations is to make  first dredge-up occur deeper in the low-metallicity
stars but only
low-metallicity stars in M92 and M15.

\subsection{The inconsistency at low metallicity}
We are left wondering why there is such an inconsistent picture of globular
clusters at low metallicity. M92 and M15 possess large spreads in [C/Fe] whilst
NGC 5466 shows a much narrower distribution.  We find that in the cluster
where we have uniform observations thermohaline mixing appears to model the
depletion of carbon well. We have already discussed the possibility that
combining studies in M15 may be effecting our interpretation of the data. 
Although data for M92 has been homogenized it is still the amalgamation of three
different studies. Compiling data from various sources may
lead to
systematic errors. Systematic offsets of 0.3 dex in [C/Fe] are possible and
could be the cause of the apparent
contradiction.
Because we have two clusters at the same metallicity in M92 and
M15 displaying similar
behavior one has to wonder how
likely it is that systematic offsets conspire in such a way to appear to give
the same effect. 

We note that the
clusters showing the largest discrepancy with
standard models, are also the most massive. Having measured the
integrated V magnitudes of the three metal-poor clusters
\citet{1997ApJ...474..223G} find
that NGC 5466 ($\approx 1.33  \times 10^5 \ \Mo$) is a lower
mass system than M92 ($\approx 3.64  \times 10^5 \ \Mo$) and M15 ($\approx
9.84  \times 10^5 \ \Mo$).
This raises the question whether there is a correlation between cluster mass and
the extent of primordial [C/Fe] spread at low metallicitites.
        The clusters M15 and M92 may have been able to sustain a greater degree
        of primordial inhomogeneous carbon enrichment than NGC 5466, with a
        resulting greater dispersion in [C/Fe] at all magnitudes on the main
        sequence and red giant branch than in NGC 5466. The greater range in
        [C/Fe] along the RGB of M15 and M92 can contribute to a greater 
        observational uncertainty in identifying the magnitude at which carbon
        depletions produced by extra mixing set in. By contrast, a smaller
        primordial carbon spread in NGC 5466 produces a more tightly defined
        locus of [C/Fe] versus $M_{\rm{V}}$ on the RGB of that cluster, making
it a
        preferable test case for mixing studies. When comparing the r-process
        elements both M92 and M15 show similar star-to-star variations
        \citep{2011ApJ...732L..17R, 2011AJ....142...22R} (although this has recently been questioned by \citet{2011ApJ...740L..38C}) which may add
additional support
        to the Na and O abundance evidence that both clusters have sustained
        heterogeneous enrichment across a range of chemical elements. It would
        be interesting to determine the O, Na, and r-process element patterns 
        in NGC 5466 to see if they are more homogeneous than in M92 and M15,
        as a mass-dependent primordial enrichment scenario would anticipate.

\subsection{The role of thermohaline mixing} \label{rolethm}
It is of course possible that thermohaline mixing does not
govern the surface composition of low mass giants. It is not
the only process by which mixing may occur in the
radiative zone. Mechanisms such as those listed in Section \ref{intro} may also
be involved.  The way in
which these
mechanisms interact is
uncertain and as they are inherently 
three-dimensional our understanding of their behavior will improve with the
study of these processes in hydrodynamical codes
\citep{2001AAS...198.6513D,2001APS..DCM.R1047B,2002ASPC..259...72T,
2002AAS...20114409E,2003ASPC..293....1B}. Still, the current generation of 1D
codes are the precursors to more sophisticated modelling and much insight can be
gained through them. One-dimensional
spherically-symmetric models with multiple 
extra-mixing mechanisms  include those of \citet{2010A&A...521A...9C} and
\citet{2010A&A...522A..10C},  both of whom demonstrate that 
the thermohaline mixing diffusion coefficient is larger than the radial
component of rotational mixing. The former also show the 
thermohaline coefficient to be larger than that of magnetic buoyancy. These
codes do not explicitly treat the interaction of the 
mixing mechanisms and how any given instability may react due to the presence of
other processes; rather, they simply add the diffusion co-efficients. \citet{2007A&A...476L..29C} 
have investigated an instance of multi-process interaction through linear
analysis and argue that
magnetic fields could serve to inhibit the effects 
of thermohaline mixing.

\citet{2010A&A...522A..10C} and \citet{2011arXiv1109.5704L} have produced detailed rotational-thermohaline
mixing models for stars of various mass and metallicity. They find
rotation leads to a deeper penetration of the convective envelope than their
non-rotating models. This allows rotational models to
begin
thermohaline
mixing at fainter magnitudes than their static models; this may go some way to
reconciling the magnitude at which stars in M92 and M15 begin mixing. It should
be
noted that low-mass, low-metallicity rotational models have been
produced by
\citet{2006A&A...453..261P} and \citet{2006ApJ...641.1087D} and in the former,
the depth of first dredge-up depends on the rotational model used. A window of
0.13 magnitudes is possible for the location of the bump depending on the treatment
of rotational physics. That is, different implementations of rotation can move
the location of the LF bump to higher or lower magnitudes.  Different models of
rotational mixing are
at play here, in much the same way that different codes produce different
third dredge-up results based on the treatment of convective stability
\citep{2005ARA&A..43..435H}. However
stochastic variations such as rotation, may lead to deeper first dredge-up
in the stars of M15 and M92. \citet{1983ApJ...275..737P} have shown that stars
on the horizontal branch of M13
rotate twice as fast as those on the horizontal branch of M3. Although we find
no major discrepancies between the location of the LF bump in these two
clusters (0.07 magnitudes \citealt{2011arXiv1109.2118N}), in the case of M15 and M92 the effect of rotation on the depth of the
FDU may be more pronounced. Why the stars in M13 would rotate faster than the
stars in M3 would also require an explanation.

Whilst it is generally agreed that the burning of \el{3}{He} in a homogenized
zone causes an instability, the efficiency of the resultant mixing  has been the
cause for debate. The
exact value of \textit{C$_{t}$} to adopt remains contentious. Recent
multi-dimensional models of thermohaline mixing by \citet{2010ApJ...723..563D},
\citet{2011ApJ...727L...8D} and \citet{2011ApJ...728L..29T} support the view of
\citet{2010A&A...521A...9C} and the suggestion by \citet{1980A&A....91..175K}
that the ``aspect ratio", $\alpha$, of the rising element should take a value
of
$\alpha \approx 1$ (\textit{C$_{t}$}$ \approx 12$). Laboratory experiments by
Stommel and Faller published in \citet{stern} have lead
\citet{1972ApJ...172..165U} and \citet{2007A&A...467L..15C} to prefer a value of
$\alpha$ somewhere closer to $\alpha \approx 6$ (\textit{C$_{t}$}$ \approx
1000$). Indeed there is great empirical evidence to prefer this value as it
appears to match observations of stars across a great range of mass and
metallicity.  The mechanism is an elegant means of matching C and N abundances in
various stars as well as ensuring measurements of \el{3}{He} in HII regions are
consistent with predictions from stellar models and Big Bang nucleosynthesis 
\citep{1984ApJ...280..629R,1995ApJ...441L..17H,1995ApJ...453L..41C,
1998A&A...336..915C,1998SSRv...84..207T,2000A&A...355...69P,
2003MNRAS.346..295R}.
Canonical models
predict that low-mass, main-sequence stars are net producers of \el{3}{He} which
is returned to the ISM through mass loss. 
\citet{1995PhRvL..75.3977H} have shown that about 90\% of the \el{3}{He} produced on the main sequence 
must be destroyed to reconcile the two fields. Our preferred value of
\textit{C$_{t}$}$= 1000$ matches the carbon and nitrogen in clusters and also
destroys over 90\% of the \el{3}{He} in the star before the tip of the RGB. The
lower value of \textit{C$_{t}$}$= 12$ once again creates a discrepancy between
measurements of \el{3}{He} in HII regions as only $\approx 20\%$ of the initial
\el{3}{He} is destroyed meaning low-mass stars are once again  producers of
\el{3}{He}. 

It is the aim of this work and in \citet{2011ApJ...728...79A} to rely on
empirical evidence to determine the best value of \textit{C$_{t}$} to use in the
1D codes. In the models presented here \textit{C$_{t}$} has been fixed at
a value of 1000. This is consistent with previous studies where the same factor
has reproduced observations of globular cluster
stars \citep{2011ApJ...728...79A}, field stars \citep{2010A&A...522A..10C} and
where the mixing has been used to explain the dichotomy between extremely metal
poor stars and carbon enhanced metal poor stars \citep{2009MNRAS.396.2313S} (although the latter point is in disagreement with \citet{2008ApJ...679.1541D, 2008ApJ...684..626D}).
\citet{2011ApJ...728...79A} demonstrate that once \textit{C$_{t}$} exceeds 1000,
the processing enters into the burning limited regime whereby increasing the
diffusion coefficient has little effect on the final surface abundances. They
also show values
of \textit{C$_{t}$}$ \ < 600$ simply lead to less mixing than is required
to match the observations.  Although we
are modelling the mixing as
\textit{diffusive} it is in fact an \textit{advective} process. 
We have implemented a linear model of thermohaline mixing which
operates via a diffusion equation. 
Recent 3D numerical simulations of thermohaline mixing 
\citep{2011ApJ...727L...8D, 2011ApJ...728L..29T} find blob-like structures, which are identified with the aspect ratio of the idealized ``fingers" in the 1D derivation of \citet{1972ApJ...172..165U}. It is from this identification that the preference for low values of \textit{C$_{t}$} in the numerical simulations 
is based. Numerical simulations are a very powerful tool, but are subject to numerous caveats - resolution (time and space), viscosity (numerical and molecular), and extrapolation to stellar conditions  (e.g., Prandtl numbers that are $\approx 10^5$ too large). In view of these uncertainties we feel it premature to draw definite conclusions from the numerical simulations. Nonetheless, if the low values of \textit{C$_{t}$} coming from the simulations prove to hold then it would result in a conclusion that either the diffusive approximation to thermohaline mixing used here is not applicable or that thermohaline mixing is too weak to produce the observed carbon depletions of globular cluster red giants.

\section{Conclusions}
We have modelled the depletion of carbon via thermohaline mixing on the RGB of
the intermediate metallicity clusters M3 and M13 as well as the metal poor
clusters NGC 5466, M92 and M15.  We conclude:

\begin{itemize}

\item Thermohaline mixing in the presence of primordial enrichment can account
for
the carbon variation seen in M3 and M13. The spread of carbon along the RGB in
both
clusters can be covered by the same set of models.
In order to match
the main-sequence spread in M13, we require a model further depleted in [C/Fe].
This model however, predicts that some stars on the RGB  should have
[C/Fe] values lower than seen. 
The majority of stars in M13 are enriched in nitrogen
whilst about half the stars in M3 appear
to be CN strong.
The results from the hybrid picture here are consistent with the
findings of \citet{2008MNRAS.390..693D} and their requirements to match the
 horizontal branch morphologies of the clusters. In their work the enriched
\el{4}{He} of the second population acts as the second parameter and helps
dictate where the star falls along the horizontal branch. 
The stars in M13 that are most enriched in [N/Fe] suggest they 
were formed with some primary
nitrogen (higher C+N+O than those in M3).

\item Thermohaline mixing can explain carbon depletion with magnitude in NGC
5466.
A single solar scaled model is sufficient to explain the cluster even though
a modest CN spread has been determined \citep{2010AJ....140.1119S}. Our models
provide 
a good fit
for this cluster if we adopt the LF bump determined by
\citet{2008AJ....136.2522M}. However, the data do not exclude the possibility
NGC 5466 is depleting carbon before the LF bump as seems to be the case for M15 and M92.

\item In M92 and M15 we have combined observations of [C/Fe] from multiple
studies.
Thermohaline mixing is unable to reproduce the evolution of carbon along the
giant branch in these clusters. The models do not deplete carbon rapidly enough along the RGB. In addition depletion appears to begin before first
dredge-up in M92,
although
this may be due to the fact we only have a few observations at low luminosity. A
targeted study at low luminosity is required to confirm if the cluster has the
same spread in [C/Fe] as M15. Both clusters appear to be mixing before the LF
bump and bright stars are not observed near the predicted carbon abundances for
initial [C/Fe]=0. 

\item  Both M92 and M15 seem to require deeper FDU in order for
the models to fit the current observations but this is not the case for NGC
5466.

\item A consideration that inevitably has to serve as a caveat to our 
        discussions is that the comparison between the thermohaline-mixing 
        models and the [C/Fe] as a function of magnitude data may be hindered by
the 
        heterogeneity of the latter, particularly in clusters such as M92 and 
        M15 where zero-point offsets of up to 0.3 dex may exist between
        the various data sources compiled. Although attempts have been made to
        compensate for these, it remains nonetheless a source of concern.  
        Determining the behavior of carbon to faint luminosities on the RGB in
        both M92 and M15 as well as M3, using a single spectrograph and data 
        analysis system could provide a more homogeneous and rigorous test of 
        our models. Such data would need to include C and N values for subgiants also. In addition, observations of lithium abundances over a large
luminosity range can be used to better define the character of extra mixing,
while measurements
of nitrogen, oxygen and/or sodium abundances can define the patterns of
primordial enrichment within the clusters.

\end{itemize}

\section{Acknowledgements}
We thank Matt Shetrone and Sarah Martell for discussions with us at Nuclei
in the
Cosmos XI. We also thank Sandro Chieffi and Marco Limongi for comparing their
models with us and demonstrating in detail the effects of rotation. We also
thank Sandro for discussions on the LF bump in low luminosity clusters and Judy Cohen for her help with the M13 data. 
GCA acknowledges the financial support of the APA scholarship.
RPC is funded by a Marie-Curie Intra-European fellowship, grant number
252431, under the European commission's FP7 framework. RJS  is a
Stromlo fellow and during his time at Monash, he was funded by the Australian
Research Council's Discovery Projects Scheme under grant DP0879472. This work
was 
supported by the NCI National Facility at the ANU.

%%%%%%%%%%%   BIBLIOGRAPHY  %%%%%%%%%%%%%%%%

\bibliographystyle{apj}

\begin{thebibliography}{141}
\expandafter\ifx\csname natexlab\endcsname\relax\def\natexlab#1{#1}\fi

\bibitem[{{Alves} {et~al.}(2004){Alves}, {Cook}, \&
  {Wishnow}}]{2004AAS...205.5301A}
{Alves}, D.~R., {Cook}, K.~H., \& {Wishnow}, E. 2004, in Bulletin of the
  American Astronomical Society, Vol.~36, Bulletin of the American Astronomical
  Society, 1425--+

\bibitem[{{Anderson} {et~al.}(2009){Anderson}, {Piotto}, {King}, {Bedin}, \&
  {Guhathakurta}}]{2009ApJ...697L..58A}
{Anderson}, J., {Piotto}, G., {King}, I.~R., {Bedin}, L.~R., \& {Guhathakurta},
  P. 2009, \apjl, 697, L58

\bibitem[{{Angelou} {et~al.}(2011){Angelou}, {Church}, {Stancliffe},
  {Lattanzio}, \& {Smith}}]{2011ApJ...728...79A}
{Angelou}, G.~C., {Church}, R.~P., {Stancliffe}, R.~J., {Lattanzio}, J.~C., \&
  {Smith}, G.~H. 2011, \apj, 728, 79

\bibitem[{{Angelou} {et~al.}(2010){Angelou}, {Lattanzio}, {Church}, \&
  {Stancliffe}}]{2010MmSAI..81.1057A}
{Angelou}, G.~C., {Lattanzio}, J.~C., {Church}, R.~P., \& {Stancliffe}, R.~J.
  2010, \memsai, 81, 1057

\bibitem[{{Arp}(1955)}]{1955AJ.....60..317A}
{Arp}, H.~C. 1955, \aj, 60, 317

\bibitem[{{Arp} {et~al.}(1952){Arp}, {Baum}, \&
  {Sandage}}]{1952AJ.....57....4A}
{Arp}, H.~C., {Baum}, W.~A., \& {Sandage}, A.~R. 1952, \aj, 57, 4

\bibitem[{{Bazan} {et~al.}(2001){Bazan}, {Castor}, {Dearborn}, {Dossa},
  {Eastman}, \& {Taylor}}]{2001APS..DCM.R1047B}
{Bazan}, G., {Castor}, J., {Dearborn}, D.~S.~P., {Dossa}, D., {Eastman}, R., \&
  {Taylor}, A. 2001, APS Meeting Abstracts, 1047

\bibitem[{{Baz{\'a}n} {et~al.}(2003){Baz{\'a}n}, {Dearborn}, {Dossa},
  {Eggleton}, {Taylor}, {Castor}, {Murray}, {Cook}, {Eltgroth}, {Cavallo},
  {Turcotte}, {Keller}, \& {Pudliner}}]{2003ASPC..293....1B}
{Baz{\'a}n}, G., {et~al.} 2003, in Astronomical Society of the Pacific
  Conference Series, Vol. 293, 3D Stellar Evolution, ed. {S.~Turcotte,
  S.~C.~Keller, \& R.~M.~Cavallo}, 1--58381

\bibitem[{{Bellman} {et~al.}(2001){Bellman}, {Briley}, {Smith}, \&
  {Claver}}]{2001PASP..113..326B}
{Bellman}, S., {Briley}, M.~M., {Smith}, G.~H., \& {Claver}, C.~F. 2001, \pasp,
  113, 326

\bibitem[{{Briley} \& {Cohen}(2001)}]{2001AJ....122..242B}
{Briley}, M.~M., \& {Cohen}, J.~G. 2001, \aj, 122, 242

\bibitem[{{Briley} {et~al.}(2002){Briley}, {Cohen}, \&
  {Stetson}}]{2002ApJ...579L..17B}
{Briley}, M.~M., {Cohen}, J.~G., \& {Stetson}, P.~B. 2002, \apjl, 579, L17

\bibitem[{{Briley} {et~al.}(2004){Briley}, {Cohen}, \&
  {Stetson}}]{2004AJ....127.1579B}
---. 2004, \aj, 127, 1579

\bibitem[{{Briley} {et~al.}(1999){Briley}, {Grundahl}, \&
  {Andersen}}]{1999AAS...195.0303B}
{Briley}, M.~M., {Grundahl}, F., \& {Andersen}, M.~I. 1999, 31, 1369

\bibitem[{{Briley} {et~al.}(1991){Briley}, {Hesser}, \&
  {Bell}}]{1991ApJ...373..482B}
{Briley}, M.~M., {Hesser}, J.~E., \& {Bell}, R.~A. 1991, \apj, 373, 482

\bibitem[{{Brown} \& {Wallerstein}(1993)}]{1993AJ....106..133B}
{Brown}, J.~A., \& {Wallerstein}, G. 1993, \aj, 106, 133

\bibitem[{{Buckley} \& {Longmore}(1992)}]{1992MNRAS.257..731B}
{Buckley}, D.~R.~V., \& {Longmore}, A.~J. 1992, \mnras, 257, 731

\bibitem[{{Buonanno} {et~al.}(1994){Buonanno}, {Corsi}, {Buzzoni}, {Cacciari},
  {Ferraro}, \& {Fusi Pecci}}]{1994A&A...290...69B}
{Buonanno}, R., {Corsi}, C.~E., {Buzzoni}, A., {Cacciari}, C., {Ferraro},
  F.~R., \& {Fusi Pecci}, F. 1994, \aap, 290, 69

\bibitem[{{Buonanno} {et~al.}(1985){Buonanno}, {Corsi}, \& {Fusi
  Pecci}}]{1985A&A...145...97B}
{Buonanno}, R., {Corsi}, C.~E., \& {Fusi Pecci}, F. 1985, \aap, 145, 97

\bibitem[{{Busso} {et~al.}(2007){Busso}, {Wasserburg}, {Nollett}, \&
  {Calandra}}]{2007ApJ...671..802B}
{Busso}, M., {Wasserburg}, G.~J., {Nollett}, K.~M., \& {Calandra}, A. 2007,
  \apj, 671, 802

\bibitem[{{Campbell} \& {Lattanzio}(2008)}]{2008A&A...490..769C}
{Campbell}, S.~W., \& {Lattanzio}, J.~C. 2008, \aap, 490, 769

\bibitem[{{Cannon} {et~al.}(1998){Cannon}, {Croke}, {Bell}, {Hesser}, \&
  {Stathakis}}]{1998MNRAS.298..601C}
{Cannon}, R.~D., {Croke}, B.~F.~W., {Bell}, R.~A., {Hesser}, J.~E., \&
  {Stathakis}, R.~A. 1998, \mnras, 298, 601

\bibitem[{{Cantiello} \& {Langer}(2010)}]{2010A&A...521A...9C}
{Cantiello}, M., \& {Langer}, N. 2010, \aap, 521, A9+

\bibitem[{{Carbon} {et~al.}(1982){Carbon}, {Romanishin}, {Langer}, {Butler},
  {Kemper}, {Trefzger}, {Kraft}, \& {Suntzeff}}]{1982ApJS...49..207C}
{Carbon}, D.~F., {Romanishin}, W., {Langer}, G.~E., {Butler}, D., {Kemper}, E.,
  {Trefzger}, C.~F., {Kraft}, R.~P., \& {Suntzeff}, N.~B. 1982, \apjs, 49, 207

\bibitem[{{Catelan} {et~al.}(2002){Catelan}, {Borissova}, {Ferraro}, {Corwin},
  {Smith}, \& {Kurtev}}]{2002AJ....124..364C}
{Catelan}, M., {Borissova}, J., {Ferraro}, F.~R., {Corwin}, T.~M., {Smith},
  H.~A., \& {Kurtev}, R. 2002, \aj, 124, 364

\bibitem[{{Chaboyer} {et~al.}(1992){Chaboyer}, {Sarajedini}, \&
  {Demarque}}]{1992ApJ...394..515C}
{Chaboyer}, B., {Sarajedini}, A., \& {Demarque}, P. 1992, \apj, 394, 515

\bibitem[{{Chanam{\'e}} {et~al.}(2005){Chanam{\'e}}, {Pinsonneault}, \&
  {Terndrup}}]{2005ApJ...631..540C}
{Chanam{\'e}}, J., {Pinsonneault}, M., \& {Terndrup}, D.~M. 2005, \apj, 631,
  540

\bibitem[{{Charbonnel}(1995)}]{1995ApJ...453L..41C}
{Charbonnel}, C. 1995, \apjl, 453, L41+

\bibitem[{{Charbonnel}(2010)}]{2010IAUS..266..131C}
---. 2010, 266, 131

\bibitem[{{Charbonnel} {et~al.}(1998){Charbonnel}, {Brown}, \&
  {Wallerstein}}]{1998A&A...332..204Cgr}
{Charbonnel}, C., {Brown}, J.~A., \& {Wallerstein}, G. 1998, \aap, 332, 204

\bibitem[{{Charbonnel} \& {Do Nascimento}(1998)}]{1998A&A...336..915C}
{Charbonnel}, C., \& {Do Nascimento}, Jr., J.~D. 1998, \aap, 336, 915

\bibitem[{{Charbonnel} \& {Lagarde}(2010)}]{2010A&A...522A..10C}
{Charbonnel}, C., \& {Lagarde}, N. 2010, \aap, 522, A10+

\bibitem[{{Charbonnel} \& {Zahn}(2007{\natexlab{b}})}]{2007A&A...476L..29C}
{Charbonnel}, C., \& {Zahn}, J. 2007{\natexlab{b}}, \aap, 476, L29

\bibitem[{{Charbonnel} \& {Zahn}(2007{\natexlab{a}})}]{2007A&A...467L..15C}
---. 2007{\natexlab{b}}, \aap, 467, L15

\bibitem[{{Cohen}(1978)}]{1978ApJ...223..487C}
{Cohen}, J.~G. 1978, \apj, 223, 487

\bibitem[{{Cohen}(1999)}]{1999AJ....117.2428C}
---. 1999, \aj, 117, 2428

\bibitem[{{Cohen}(2011)}]{2011ApJ...740L..38C}
---. 2011, \apjl, 740, L38

\bibitem[{{Cohen} {et~al.}(2005){Cohen}, {Briley}, \&
  {Stetson}}]{2005AJ....130.1177C}
{Cohen}, J.~G., {Briley}, M.~M., \& {Stetson}, P.~B. 2005, \aj, 130, 1177

\bibitem[{{Cohen} \& {Mel{\'e}ndez}(2005)}]{2005AJ....129..303C}
{Cohen}, J.~G., \& {Mel{\'e}ndez}, J. 2005, \aj, 129, 303

\bibitem[{{D'Antona} {et~al.}(2005){D'Antona}, {Bellazzini}, {Caloi}, {Pecci},
  {Galleti}, \& {Rood}}]{2005ApJ...631..868D}
{D'Antona}, F., {Bellazzini}, M., {Caloi}, V., {Pecci}, F.~F., {Galleti}, S.,
  \& {Rood}, R.~T. 2005, \apj, 631, 868

\bibitem[{{D'Antona} \& {Caloi}(2008)}]{2008MNRAS.390..693D}
{D'Antona}, F., \& {Caloi}, V. 2008, \mnras, 390, 693

\bibitem[{{de Mink} {et~al.}(2009){de Mink}, {Pols}, {Langer}, \&
  {Izzard}}]{2009A&A...507L...1D}
{de Mink}, S.~E., {Pols}, O.~R., {Langer}, N., \& {Izzard}, R.~G. 2009, \aap,
  507, L1

\bibitem[{{Dearborn} {et~al.}(2001){Dearborn}, {Bazan}, {Castor}, {Cavallo},
  {Cohl}, {Cook}, {Dossa}, {Eastman}, {Eggleton}, {Eltgroth}, {Keller},
  {Murray}, {Taylor}, {Turcotte}, \& {Djehuty Team}}]{2001AAS...198.6513D}
{Dearborn}, D.~S.~P., {et~al.} 2001, in Bulletin of the American Astronomical
  Society, Vol.~33, Bulletin of the American Astronomical Society, 886--+

\bibitem[{{Decressin} {et~al.}(2007){Decressin}, {Meynet}, {Charbonnel},
  {Prantzos}, \& {Ekstr{\"o}m}}]{2007A&A...464.1029D}
{Decressin}, T., {Meynet}, G., {Charbonnel}, C., {Prantzos}, N., \&
  {Ekstr{\"o}m}, S. 2007, \aap, 464, 1029


\bibitem[{{Denissenkov}(2010)}]{2010ApJ...723..563D}
{Denissenkov}, P.~A. 2010, \apj, 723, 563

\bibitem[{{Denissenkov} {et~al.}(2006){Denissenkov}, {Chaboyer}, \&
  {Li}}]{2006ApJ...641.1087D}
{Denissenkov}, P.~A., {Chaboyer}, B., \& {Li}, K. 2006, \apj, 641, 1087

\bibitem[{{Denissenkov} {et~al.}(1998){Denissenkov}, {Da Costa}, {Norris}, \&
  {Weiss}}]{1998A&A...333..926D}
{Denissenkov}, P.~A., {Da Costa}, G.~S., {Norris}, J.~E., \& {Weiss}, A. 1998,
  \aap, 333, 926

\bibitem[{{Denissenkov} \& {Merryfield}(2011)}]{2011ApJ...727L...8D}
{Denissenkov}, P.~A., \& {Merryfield}, W.~J. 2011, \apjl, 727, L8+

\bibitem[{{Denissenkov} \& {Pinsonneault}(2008{\natexlab{a}})}]{2008ApJ...679.1541D}
{Denissenkov}, P.~A., \& {Pinsonneault}, M. 2008{\natexlab{a}}, \apj, 679, 1541

\bibitem[{{Denissenkov} \& {Pinsonneault}(2008{\natexlab{b}})}]{2008ApJ...684..626D}
{Denissenkov}, P.~A., \& {Pinsonneault}, M. 2008{\natexlab{b}}, \apj, 684, 626

\bibitem[{{Denissenkov} {et~al.}(2009){Denissenkov}, {Pinsonneault}, \&
  {MacGregor}}]{2009ApJ...696.1823D}
{Denissenkov}, P.~A., {Pinsonneault}, M., \& {MacGregor}, K.~B. 2009, \apj,
  696, 1823

\bibitem[{{Denissenkov} \& {Tout}(2000)}]{2000MNRAS.316..395D}
{Denissenkov}, P.~A., \& {Tout}, C.~A. 2000, \mnras, 316, 395

\bibitem[{{Denissenkov} \& {VandenBerg}(2003)}]{2003ApJ...593..509D}
{Denissenkov}, P.~A., \& {VandenBerg}, D.~A. 2003, \apj, 593, 509

\bibitem[{{di Cecco} {et~al.}(2010){di Cecco}, {Becucci}, {Bono}, {Monelli},
  {Stetson}, {Degl'Innocenti}, {Prada Moroni}, {Nonino}, {Weiss}, {Buonanno},
  {Calamida}, {Caputo}, {Corsi}, {Ferraro}, {Iannicola}, {Pulone},
  {Romaniello}, \& {Walker}}]{2010PASP..122..991D}
{di Cecco}, A., {et~al.} 2010, \pasp, 122, 991

\bibitem[{{Eggleton} {et~al.}(2002){Eggleton}, {Bazan}, {Cavallo}, {Dearborn},
  {Dossa}, {Keller}, {Taylor}, \& {Turcotte}}]{2002AAS...20114409E}
{Eggleton}, P.~P., {Bazan}, G., {Cavallo}, R.~M., {Dearborn}, D.~S.~P.,
  {Dossa}, D.~D., {Keller}, S., {Taylor}, A.~G., \& {Turcotte}, S. 2002, in
  Bulletin of the American Astronomical Society, Vol.~35, Bulletin of the
  American Astronomical Society, 570--+

\bibitem[{{Eggleton} {et~al.}(2006){Eggleton}, {Dearborn}, \&
  {Lattanzio}}]{2006Sci...314.1580E}
{Eggleton}, P.~P., {Dearborn}, D.~S.~P., \& {Lattanzio}, J.~C. 2006, Science,
  314, 1580

\bibitem[{{Eggleton} {et~al.}(2008){Eggleton}, {Dearborn}, \&
  {Lattanzio}}]{2008ApJ...677..581E}
---. 2008, \apj, 677, 581

\bibitem[{{Fekadu} {et~al.}(2007){Fekadu}, {Sandquist}, \&
  {Bolte}}]{2007ApJ...663..277F}
{Fekadu}, N., {Sandquist}, E.~L., \& {Bolte}, M. 2007, \apj, 663, 277

\bibitem[{{Fenner} {et~al.}(2004){Fenner}, {Campbell}, {Karakas}, {Lattanzio},
  \& {Gibson}}]{2004MNRAS.353..789F}
{Fenner}, Y., {Campbell}, S., {Karakas}, A.~I., {Lattanzio}, J.~C., \&
  {Gibson}, B.~K. 2004, \mnras, 353, 789

\bibitem[{{Fusi Pecci} {et~al.}(1990){Fusi Pecci}, {Ferraro}, {Crocker},
  {Rood}, \& {Buonanno}}]{1990A&A...238...95F}
{Fusi Pecci}, F., {Ferraro}, F.~R., {Crocker}, D.~A., {Rood}, R.~T., \&
  {Buonanno}, R. 1990, \aap, 238, 95

\bibitem[{{Gnedin} \& {Ostriker}(1997)}]{1997ApJ...474..223G}
{Gnedin}, O.~Y., \& {Ostriker}, J.~P. 1997, \apj, 474, 223

\bibitem[{{Gratton} {et~al.}(2004){Gratton}, {Sneden}, \&
  {Carretta}}]{2004ARA&A..42..385G}
{Gratton}, R., {Sneden}, C., \& {Carretta}, E. 2004, \araa, 42, 385

\bibitem[{{Gratton} {et~al.}(1997){Gratton}, {Fusi Pecci}, {Carretta},
  {Clementini}, {Corsi}, \& {Lattanzi}}]{1997ApJ...491..749G}
{Gratton}, R.~G., {Fusi Pecci}, F., {Carretta}, E., {Clementini}, G., {Corsi},
  C.~E., \& {Lattanzi}, M. 1997, \apj, 491, 749

\bibitem[{{Gratton} {et~al.}(2000){Gratton}, {Sneden}, {Carretta}, \&
  {Bragaglia}}]{2000A&A...354..169G}
{Gratton}, R.~G., {Sneden}, C., {Carretta}, E., \& {Bragaglia}, A. 2000, \aap,
  354, 169

\bibitem[{{Grundahl} {et~al.}(2000){Grundahl}, {VandenBerg}, {Bell},
  {Andersen}, \& {Stetson}}]{2000AJ....120.1884G}
{Grundahl}, F., {VandenBerg}, D.~A., {Bell}, R.~A., {Andersen}, M.~I., \&
  {Stetson}, P.~B. 2000, \aj, 120, 1884

\bibitem[{{Harris}(1996)}]{1996AJ....112.1487H}
{Harris}, W.~E. 1996, \aj, 112, 1487

\bibitem[{{Hata} {et~al.}(1995){Hata}, {Scherrer}, {Steigman}, {Thomas},
  {Walker}, {Bludman}, \& {Langacker}}]{1995PhRvL..75.3977H}
{Hata}, N., {Scherrer}, R.~J., {Steigman}, G., {Thomas}, D., {Walker}, T.~P.,
  {Bludman}, S., \& {Langacker}, P. 1995, Physical Review Letters, 75, 3977

\bibitem[{{Herwig}(2005)}]{2005ARA&A..43..435H}
{Herwig}, F. 2005, \araa, 43, 435

\bibitem[{{Hesser} \& {Bell}(1980)}]{1980ApJ...238L.149H}
{Hesser}, J.~E., \& {Bell}, R.~A. 1980, \apjl, 238, L149

\bibitem[{{Hogan}(1995)}]{1995ApJ...441L..17H}
{Hogan}, C.~J. 1995, \apjl, 441, L17

\bibitem[{{Hubbard} \& {Dearborn}(1980)}]{1980ApJ...239..248H}
{Hubbard}, E.~N., \& {Dearborn}, D.~S.~P. 1980, \apj, 239, 248

\bibitem[{{Iben}(1967)}]{1967ApJ...147..624I}
{Iben}, Jr., I. 1967, \apj, 147, 624

\bibitem[{{Iben} \& {Ehrman}(1962)}]{1962ApJ...135..770I}
{Iben}, Jr., I., \& {Ehrman}, J.~R. 1962, \apj, 135, 770

\bibitem[{{Iben} \& {Faulkner}(1968)}]{1968ApJ...153..101I}
{Iben}, Jr., I., \& {Faulkner}, J. 1968, \apj, 153, 101

\bibitem[{{Jimenez} {et~al.}(1996){Jimenez}, {Thejll}, {Jorgensen},
  {MacDonald}, \& {Pagel}}]{1996MNRAS.282..926J}
{Jimenez}, R., {Thejll}, P., {Jorgensen}, U.~G., {MacDonald}, J., \& {Pagel},
  B. 1996, \mnras, 282, 926

\bibitem[{{Johnson} \& {Sandage}(1955)}]{1955ApJ...121..616J}
{Johnson}, H.~L., \& {Sandage}, A.~R. 1955, \apj, 121, 616

\bibitem[{{Kippenhahn} {et~al.}(1980){Kippenhahn}, {Ruschenplatt}, \&
  {Thomas}}]{1980A&A....91..175K}
{Kippenhahn}, R., {Ruschenplatt}, G., \& {Thomas}, H. 1980, \aap, 91, 175


\bibitem[{{Kraft}(1994)}]{1994PASP..106..553K}
{Kraft}, R.~P.  1994, \pasp, 106, 553

\bibitem[{{Kraft} {et~al.}(1992){Kraft}, {Sneden}, {Langer}, \&
  {Prosser}}]{1992AJ....104..645K}
{Kraft}, R.~P., {Sneden}, C., {Langer}, G.~E., \& {Prosser}, C.~F. 1992, \aj,
  104, 645

\bibitem[{{Lagarde} {et~al.}(2011){Lagarde}, {Charbonnel}, {Decressin}, \&
  {Hagelberg}}]{2011arXiv1109.5704L}
{Lagarde}, N., {Charbonnel}, C., {Decressin}, T., \& {Hagelberg}, J. 2011, ArXiv:1109.5704 

\bibitem[{{Langer} {et~al.}(1986){Langer}, {Kraft}, {Carbon}, {Friel}, \&
  {Oke}}]{1986PASP...98..473L}
{Langer}, G.~E., {Kraft}, R.~P., {Carbon}, D.~F., {Friel}, E., \& {Oke}, J.~B.
  1986, \pasp, 98, 473

\bibitem[{{Lee}(1999)}]{1999AJ....118..920L}
{Lee}, S. 1999, \aj, 118, 920

\bibitem[{{Lee} {et~al.}(2005){Lee}, {Joo}, {Han}, {Chung}, {Ree}, {Sohn},
  {Kim}, {Yoon}, {Yi}, \& {Demarque}}]{2005ApJ...621L..57L}
{Lee}, Y., {et~al.} 2005, \apjl, 621, L57

\bibitem[{{Lind} {et~al.}(2009){Lind}, {Primas}, {Charbonnel}, {Grundahl}, \&
  {Asplund}}]{2009A&A...503..545L}
{Lind}, K., {Primas}, F., {Charbonnel}, C., {Grundahl}, F., \& {Asplund}, M.
  2009, \aap, 503, 545

\bibitem[{{Martell} {et~al.}(2008{\natexlab{a}}){Martell}, {Smith}, \&
  {Briley}}]{2008PASP..120....7M}
{Martell}, S.~L., {Smith}, G.~H., \& {Briley}, M.~M. 2008{\natexlab{a}}, \pasp,
  120, 7

\bibitem[{{Martell} {et~al.}(2008{\natexlab{b}}){Martell}, {Smith}, \&
  {Briley}}]{2008AJ....136.2522M}
---. 2008{\natexlab{b}}, \aj, 136, 2522

\bibitem[{{Mucciarelli} {et~al.}(2011){Mucciarelli}, {Salaris}, {Lovisi},
  {Ferraro}, {Lanzoni}, {Lucatello}, \& {Gratton}}]{2011MNRAS.412...81M}
{Mucciarelli}, A., {Salaris}, M., {Lovisi}, L., {Ferraro}, F.~R., {Lanzoni},
  B., {Lucatello}, S., \& {Gratton}, R.~G. 2011, \mnras, 412, 81
  
  \bibitem[{{Nataf} {et~al.}(2011){Nataf}, {Gould}, {Pinsonneault}, \&
  {Udalski}}]{2011arXiv1109.2118N}
{Nataf}, D.~M., {Gould}, A.~P., {Pinsonneault}, M.~H., \& {Udalski}, A. 2011,
ArXiv:1109.2118

\bibitem[{{Nordhaus} {et~al.}(2008){Nordhaus}, {Busso}, {Wasserburg},
  {Blackman}, \& {Palmerini}}]{2008ApJ...684L..29N}
{Nordhaus}, J., {Busso}, M., {Wasserburg}, G.~J., {Blackman}, E.~G., \&
  {Palmerini}, S. 2008, \apjl, 684, L29

\bibitem[{{Norris} {et~al.}(1981){Norris}, {Cottrell}, {Freeman}, \& {Da
  Costa}}]{1981ApJ...244..205N}
{Norris}, J., {Cottrell}, P.~L., {Freeman}, K.~C., \& {Da Costa}, G.~S. 1981,
  \apj, 244, 205

\bibitem[{{Norris} \& {Pilachowski}(1985)}]{1985ApJ...299..295N}
{Norris}, J., \& {Pilachowski}, C.~A. 1985, \apj, 299, 295

\bibitem[{{Norris} \& {Smith}(1984)}]{1984ApJ...287..255N}
{Norris}, J., \& {Smith}, G.~H. 1984, \apj, 287, 255

\bibitem[{{Oke} \& {Schwarzschild}(1952)}]{1952ApJ...116..317O}
{Oke}, J.~B., \& {Schwarzschild}, M. 1952, \apj, 116, 317

\bibitem[{{Osborn}(1971)}]{1971Obs....91..223O}
{Osborn}, W. 1971, The Observatory, 91, 223

\bibitem[{{Palacios} {et~al.}(2006){Palacios}, {Charbonnel}, {Talon}, \&
  {Siess}}]{2006A&A...453..261P}
{Palacios}, A., {Charbonnel}, C., {Talon}, S., \& {Siess}, L. 2006, \aap, 453,
  261

\bibitem[{{Palla} {et~al.}(2000){Palla}, {Bachiller}, {Stanghellini}, {Tosi},
  \& {Galli}}]{2000A&A...355...69P}
{Palla}, F., {Bachiller}, R., {Stanghellini}, L., {Tosi}, M., \& {Galli}, D.
  2000, \aap, 355, 69

\bibitem[{{Palmerini} {et~al.}(2011{\natexlab{b}}){Palmerini}, {Cristallo},
  {Busso}, {Abia}, {Uttenthaler}, {Gialanella}, \&
  {Maiorca}}]{2011ApJ...741...26P}
{Palmerini}, S., {Cristallo}, S., {Busso}, M., {Abia}, C., {Uttenthaler}, S.,
  {Gialanella}, L., \& {Maiorca}, E. 2011{\natexlab{b}}, \apj, 741, 26

\bibitem[{{Palmerini} {et~al.}(2011{\natexlab{a}}){Palmerini}, {La Cognata},
  {Cristallo}, \& {Busso}}]{2011ApJ...729....3P}
{Palmerini}, S., {La Cognata}, M., {Cristallo}, S., \& {Busso}, M.
  2011{\natexlab{a}}, \apj, 729, 3


\bibitem[{{Palmerini} {et~al.}(2009){Palmerini}, {Busso}, {Maiorca}, \&
  {Guandalini}}]{2009PASA...26..161P}
{Palmerini}, S., {Busso}, M., {Maiorca}, E., \& {Guandalini}, R. 2009,
  Publications of the Astronomical Society of Australia, 26, 161

\bibitem[{{Pancino} {et~al.}(2010){Pancino}, {Rejkuba}, {Zoccali}, \&
  {Carrera}}]{2010A&A...524A..44P}
{Pancino}, E., {Rejkuba}, M., {Zoccali}, M., \& {Carrera}, R. 2010, \aap, 524,
  A44+

\bibitem[{{Paust} {et~al.}(2007){Paust}, {Chaboyer}, \&
  {Sarajedini}}]{2007AJ....133.2787P}
{Paust}, N.~E.~Q., {Chaboyer}, B., \& {Sarajedini}, A. 2007, \aj, 133, 2787

\bibitem[{{Pavlenko} {et~al.}(2003){Pavlenko}, {Jones}, \&
  {Longmore}}]{2003MNRAS.345..311P}
{Pavlenko}, Y.~V., {Jones}, H.~R.~A., \& {Longmore}, A.~J. 2003, \mnras, 345,
  311

\bibitem[{{Peterson}(1980)}]{1980ApJ...237L..87P}
{Peterson}, R.~C. 1980, \apjl, 237, L87

\bibitem[{{Peterson}(1983)}]{1983ApJ...275..737P}
---. 1983, \apj, 275, 737

\bibitem[{{Pilachowski} {et~al.}(2003){Pilachowski}, {Sneden}, {Freeland}, \&
  {Casperson}}]{2003AJ....125..794P}
{Pilachowski}, C., {Sneden}, C., {Freeland}, E., \& {Casperson}, J. 2003, \aj,
  125, 794

\bibitem[{{Piotto}(2009)}]{2009IAUS..258..233P}
{Piotto}, G. 2009, 258, 233

\bibitem[{{Piotto} {et~al.}(2005){Piotto}, {Villanova}, {Bedin}, {Gratton},
  {Cassisi}, {Momany}, {Recio-Blanco}, {Lucatello}, {Anderson}, {King},
  {Pietrinferni}, \& {Carraro}}]{2005ApJ...621..777P}
{Piotto}, G., {et~al.} 2005, \apj, 621, 777

\bibitem[{{Piotto} {et~al.}(2007){Piotto}, {Bedin}, {Anderson}, {King},
  {Cassisi}, {Milone}, {Villanova}, {Pietrinferni}, \&
  {Renzini}}]{2007ApJ...661L..53P}
---. 2007, \apjl, 661, L53

\bibitem[{{Popper}(1947)}]{1947ApJ...105..204P}
{Popper}, D.~M. 1947, \apj, 105, 204

\bibitem[{{Recio-Blanco} \& {de Laverny}(2007)}]{2007A&A...461L..13R}
{Recio-Blanco}, A., \& {de Laverny}, P. 2007, \aap, 461, L13

\bibitem[{{Reid}(1997)}]{1997AJ....114..161R}
{Reid}, I.~N. 1997, \aj, 114, 161

\bibitem[{{Roederer}(2011)}]{2011ApJ...732L..17R}
{Roederer}, I.~U. 2011, \apjl, 732, L17+

\bibitem[{{Roederer} \& {Sneden}(2011)}]{2011AJ....142...22R}
{Roederer}, I.~U., \& {Sneden}, C. 2011, \aj, 142, 22

\bibitem[{{Romano} {et~al.}(2003){Romano}, {Tosi}, {Matteucci}, \&
  {Chiappini}}]{2003MNRAS.346..295R}
{Romano}, D., {Tosi}, M., {Matteucci}, F., \& {Chiappini}, C. 2003, \mnras,
  346, 295

\bibitem[{{Rood} {et~al.}(1984){Rood}, {Bania}, \&
  {Wilson}}]{1984ApJ...280..629R}
{Rood}, R.~T., {Bania}, T.~M., \& {Wilson}, T.~L. 1984, \apj, 280, 629

\bibitem[{{Salaris} \& {Weiss}(2002)}]{2002A&A...388..492S}
{Salaris}, M., \& {Weiss}, A. 2002, \aap, 388, 492

\bibitem[{{Sandage}(1953)}]{1953AJ.....58...61S}
{Sandage}, A.~R. 1953, \aj, 58, 61

\bibitem[{{Sandage}(1954)}]{1954MSRSL...1..254S}
---. 1954, Memoires of the Societe Royale des Sciences de Liege, 1, 254

\bibitem[{{Shetrone} {et~al.}(2010){Shetrone}, {Martell}, {Wilkerson}, {Adams},
  {Siegel}, {Smith}, \& {Bond}}]{2010AJ....140.1119S}
{Shetrone}, M., {Martell}, S.~L., {Wilkerson}, R., {Adams}, J., {Siegel},
  M.~H., {Smith}, G.~H., \& {Bond}, H.~E. 2010, \aj, 140, 1119

\bibitem[{{Shetrone}(2003)}]{2003ApJ...585L..45S}
{Shetrone}, M.~D. 2003, \apjl, 585, L45

\bibitem[{{Smith}(2002)}]{2002PASP..114.1097S}
{Smith}, G.~H. 2002, \pasp, 114, 1097

\bibitem[{{Smith}(2006)}]{2006PASP..118.1225S}
---. 2006, \pasp, 118, 1225

\bibitem[{{Smith} {et~al.}(2005){Smith}, {Briley}, \&
  {Harbeck}}]{2005AJ....129.1589S}
{Smith}, G.~H., {Briley}, M.~M., \& {Harbeck}, D. 2005, \aj, 129, 1589

\bibitem[{{Smith} \& {Martell}(2003)}]{2003PASP..115.1211S}
{Smith}, G.~H., \& {Martell}, S.~L. 2003, \pasp, 115, 1211

\bibitem[{{Smith} {et~al.}(1996){Smith}, {Shetrone}, {Bell}, {Churchill}, \&
  {Briley}}]{1996AJ....112.1511S}
{Smith}, G.~H., {Shetrone}, M.~D., {Bell}, R.~A., {Churchill}, C.~W., \&
  {Briley}, M.~M. 1996, \aj, 112, 1511

\bibitem[{{Smolinski} {et~al.}(2011){Smolinski}, {Martell}, {Beers}, \&
  {Lee}}]{2011AJ....142..126S}
{Smolinski}, J.~P., {Martell}, S.~L., {Beers}, T.~C., \& {Lee}, Y.~S. 2011,
  \aj, 142, 1268

\bibitem[{{Sneden} {et~al.}(2004){Sneden}, {Kraft}, {Guhathakurta}, {Peterson},
  \& {Fulbright}}]{2004AJ....127.2162S}
{Sneden}, C., {Kraft}, R.~P., {Guhathakurta}, P., {Peterson}, R.~C., \&
  {Fulbright}, J.~P. 2004, \aj, 127, 2162

\bibitem[{{Sneden} {et~al.}(1992){Sneden}, {Kraft}, {Prosser}, \&
  {Langer}}]{1992AJ....104.2121S}
{Sneden}, C., {Kraft}, R.~P., {Prosser}, C.~F., \& {Langer}, G.~E. 1992, \aj,
  104, 2121

\bibitem[{{Sneden} {et~al.}(1997){Sneden}, {Kraft}, {Shetrone}, {Smith},
  {Langer}, \& {Prosser}}]{1997AJ....114.1964S}
{Sneden}, C., {Kraft}, R.~P., {Shetrone}, M.~D., {Smith}, G.~H., {Langer},
  G.~E., \& {Prosser}, C.~F. 1997, \aj, 114, 1964

\bibitem[{{Sneden} {et~al.}(2000){Sneden}, {Pilachowski}, \&
  {Kraft}}]{2000AJ....120.1351S}
{Sneden}, C., {Pilachowski}, C.~A., \& {Kraft}, R.~P. 2000, \aj, 120, 1351

\bibitem[{{Sneden} {et~al.}(1986){Sneden}, {Pilachowski}, \&
  {Vandenberg}}]{1986ApJ...311..826S}
{Sneden}, C., {Pilachowski}, C.~A., \& {Vandenberg}, D.~A. 1986, \apj, 311, 826

\bibitem[{{Sollima} {et~al.}(2007){Sollima}, {Ferraro}, {Bellazzini},
  {Origlia}, {Straniero}, \& {Pancino}}]{2007ApJ...654..915S}
{Sollima}, A., {Ferraro}, F.~R., {Bellazzini}, M., {Origlia}, L., {Straniero},
  O., \& {Pancino}, E. 2007, \apj, 654, 915

\bibitem[{{Spite} {et~al.}(2005){Spite}, {Cayrel}, {Plez}, {Hill}, {Spite},
  {Depagne}, {Fran{\c c}ois}, {Bonifacio}, {Barbuy}, {Beers}, {Andersen},
  {Molaro}, {Nordstr{\"o}m}, \& {Primas}}]{2005A&A...430..655S}
{Spite}, M., {et~al.} 2005, \aap, 430, 655

\bibitem[{{Spite} {et~al.}(2006){Spite}, {Cayrel}, {Hill}, {Spite}, {Fran{\c
  c}ois}, {Plez}, {Bonifacio}, {Molaro}, {Depagne}, {Andersen}, {Barbuy},
  {Beers}, {Nordstr{\"o}m}, \& {Primas}}]{2006A&A...455..291S}
---. 2006, \aap, 455, 291


\bibitem[{{Stancliffe}(2010)}]{2010MNRAS.403..505S}
---. 2010, \mnras, 403, 505

\bibitem[{{Stancliffe} {et~al.}(2009){Stancliffe}, {Church}, {Angelou}, \&
  {Lattanzio}}]{2009MNRAS.396.2313S}
{Stancliffe}, R.~J., {Church}, R.~P., {Angelou}, G.~C., \& {Lattanzio}, J.~C.
  2009, \mnras, 396, 2313

\bibitem[{Stern(1960)}]{stern}
Stern, M. 1960, Tellus, 12

\bibitem[{{Suntzeff}(1981)}]{1981ApJS...47....1S}
{Suntzeff}, N.~B. 1981, \apjs, 47, 1

\bibitem[{{Suntzeff} \& {Smith}(1991)}]{1991ApJ...381..160S}
{Suntzeff}, N.~B., \& {Smith}, V.~V. 1991, \apj, 381, 160

\bibitem[{{Sweigart}(1978)}]{1978IAUS...80..333S}
{Sweigart}, A.~V. 1978, in IAU Symposium, Vol.~80, The HR Diagram - The 100th
  Anniversary of Henry Norris Russell, ed. {A.~G.~D.~Philip \& D.~S.~Hayes},
  333--343

\bibitem[{{Sweigart} \& {Mengel}(1979)}]{1979ApJ...229..624S}
{Sweigart}, A.~V., \& {Mengel}, J.~G. 1979, \apj, 229, 624

\bibitem[{{Tosi}(1998)}]{1998SSRv...84..207T}
{Tosi}, M. 1998, \ssr, 84, 207

\bibitem[{{Traxler} {et~al.}(2011){Traxler}, {Garaud}, \&
  {Stellmach}}]{2011ApJ...728L..29T}
{Traxler}, A., {Garaud}, P., \& {Stellmach}, S. 2011, \apjl, 728, L29+

\bibitem[{{Trefzger} {et~al.}(1983){Trefzger}, {Langer}, {Carbon}, {Suntzeff},
  \& {Kraft}}]{1983ApJ...266..144T}
{Trefzger}, D.~V., {Langer}, G.~E., {Carbon}, D.~F., {Suntzeff}, N.~B., \&
  {Kraft}, R.~P. 1983, \apj, 266, 144

\bibitem[{{Turcotte} {et~al.}(2002){Turcotte}, {Bazan}, {Castor}, {Cavallo},
  {Cohl}, {Cook}, {Dearborn}, {Dossa}, {Eastman}, {Eggleton}, {Eltgroth},
  {Keller}, {Murray}, \& {Taylor}}]{2002ASPC..259...72T}
{Turcotte}, S., {et~al.} 2002, in Astronomical Society of the Pacific
  Conference Series, Vol. 259, IAU Colloq. 185: Radial and Nonradial
  Pulsationsn as Probes of Stellar Physics, ed. {C.~Aerts, T.~R.~Bedding, \&
  J.~Christensen-Dalsgaard}, 72--+

\bibitem[{{Ulrich}(1972)}]{1972ApJ...172..165U}
{Ulrich}, R.~K. 1972, \apj, 172, 165

\bibitem[{{VandenBerg}(2000)}]{2000ApJS..129..315V}
{VandenBerg}, D.~A. 2000, \apjs, 129, 315

\bibitem[{{Ventura} \& {D'Antona}(2011)}]{2011MNRAS.410.2760V}
{Ventura}, P., \& {D'Antona}, F. 2011, \mnras, 410, 2760

\bibitem[{{Wachlin} {et~al.}(2011){Wachlin}, {Miller Bertolami}, \&
  {Althaus}}]{2011arXiv1104.0832W}
{Wachlin}, F.~C., {Miller Bertolami}, M.~M., \& {Althaus}, L.~G. 2011,
ArXiv:1104.0832

\bibitem[{{Wasserburg} {et~al.}(1995){Wasserburg}, {Boothroyd}, \&
  {Sackmann}}]{1995ApJ...447L..37W}
{Wasserburg}, G.~J., {Boothroyd}, A.~I., \& {Sackmann}, I. 1995, \apjl, 447,
  L37+

\bibitem[{{Zoccali} {et~al.}(1999){Zoccali}, {Cassisi}, {Piotto}, {Bono}, \&
  {Salaris}}]{1999ApJ...518L..49Z}
{Zoccali}, M., {Cassisi}, S., {Piotto}, G., {Bono}, G., \& {Salaris}, M. 1999,
  \apjl, 518, L49

\end{thebibliography}

\label{lastpage}

\end{document}